\documentclass[sigconf, screen, noacm]{acmart}

\usepackage{graphicx}
\usepackage{xcolor}
\usepackage{xspace}
\usepackage{fancyhdr}
\usepackage{hyperref}
\usepackage{booktabs}

\ifx\figurename\undefined \def\figurename{Figure}\fi
\renewcommand{\figurename}{Fig.}

\newcommand{\Sect}[1]{Sec.~\ref{#1}}
\newcommand{\Fig}[1]{Fig.~\ref{#1}}
\newcommand{\Tbl}[1]{Tbl.~\ref{#1}}

\newcommand{\Eqn}[1]{Eqn.~\ref{#1}}

\newcommand{\Review}[1]{}

\newcommand{\specialcell}[2][c]{\begin{tabular}[#1]{@{}c@{}}#2\end{tabular}}

\def\cF{{\mathcal{F}}}

\def\cF{{\mathcal{F}}}

\newcommand{\mode}[1]{\underline{\textsc{#1}}\xspace}

\newcommand{\proj}{\textsc{Dubot}\xspace}
\newcommand{\para}[1]{\textit{\textbf{#1}} }

\graphicspath{{figures/}}

\setcopyright{none}
\acmConference[ISCA '27]{54th IEEE/ACM International Symposium on Computer Architecture}{June 5--9, 2027}{Atlanta, Georgia, USA}
\acmYear{2027}
\acmISBN{978-x-xxxx-xxxx-x/YY/MM}
\acmDOI{XXXXXXX.XXXXXXX}
\begin{document}


\title{Resource-Aware VLA–MPC Co-Location for Efficient Embodied AI}

\author{Zeyu Guo}
\orcid{}
\affiliation{%
  \institution{Shanghai Jiao Tong University}
  \city{Shanghai}
  \country{China}
}
\email{zeyuguo77@sjtu.edu.cn}

\author{Hao Lou}
\orcid{????}
\affiliation{%
  \institution{Shanghai Jiao Tong University}
  \city{Shanghai}
  \country{China}
}
\email{louhao088@sjtu.edu.cn}

\author{Yuhui Hao}
\orcid{????}
\affiliation{%
  \institution{ICT, Chinese Academy of Sciences}
  \city{Beijing}
  \country{China}
}
\email{yuhuihao@tju.edu.cn}

\author{Han Zhao}
\authornote{Corresponding Authors.}
\orcid{????}
\affiliation{%
  \institution{Shanghai Jiao Tong University}
  \city{Shanghai}
  \country{China}
}
\email{zhao-han@cs.sjtu.edu.cn}

\author{Zihan Liu}
\orcid{0000-0002-0874-0682}
\affiliation{%
  \institution{Shanghai Jiao Tong University, Shanghai Qi Zhi Institute}
  \city{Shanghai}
  \country{China}
}
\email{altair.liu@sjtu.edu.cn}

\author{Jingwen Leng}
\orcid{0000-0002-5660-5493}
\affiliation{%
  \institution{Shanghai Jiao Tong University, Shanghai Qi Zhi Institute}
  \city{Shanghai}
  \country{China}
}
\email{leng-jw@sjtu.edu.cn}

\author{Yiming Gan}
\orcid{0000-0002-2033-5057}
\affiliation{%
  \institution{ICT, Chinese Academy of Sciences}
  \city{Beijing}
  \country{China}
}
\email{ganyiming@ict.ac.cn}

\author{Yu Feng}
\authornotemark[1]
\orcid{0000-0002-2192-5737}
\affiliation{%
  \institution{Shanghai Jiao Tong University, Shanghai Qi Zhi Institute}
  \city{Shanghai}
  \country{China}
}
\email{y-feng@sjtu.edu.cn}

\author{Minyi Guo}
\orcid{0000-0003-0034-2302}
\affiliation{%
  \institution{Shanghai Jiao Tong University, Shanghai Qi Zhi Institute}
  \city{Shanghai}
  \country{China}
}
\email{guo-my@cs.sjtu.edu.cn}

\renewcommand{\shortauthors}{Zeyu Guo et al.}


\begin{abstract}

Recent advances in embodied artificial intelligence (EAI) have unleashed new potential in many frontiers.
Current EAI systems often combine vision-language-action (VLA) inference with high-frequency model predictive control (MPC). 
On mobile GPUs, these workloads are typically serialized, causing control deadline misses. 

We present \proj, a framework that automatically schedules VLA inference and MPC on a single mobile GPU.
Specifically, \proj identifies concurrent opportunities and proposes three co-location mechanisms: spatial partitioning, block-level co-location, and intra-block fusion. 
By leveraging different mechanisms, \proj enables fine-grained co-location to improve the efficiency of various VLA-MPC operator pairs.
\proj further proposes a co-location optimizer that can determine the optimal co-location schedule under workload dependency, QoS requirements, and hardware-resource constraints. 
Across representative VLA inference paradigms on NVIDIA Jetson Orin and Thor, \proj achieves up to 1.4$\times$ end-to-end speedup and reduces QoS violations by up to 38\%.

\end{abstract}

\maketitle


\section{Introduction}
\label{sec:intro}

Embodied artificial intelligence (EAI) has rapidly moved toward real-world deployment in humanoid robots~\cite{fu2024humanplus, he2024omnih2o, he2024learning}, autonomous vehicles~\cite{hu2023_uniad, jiang2023vad, sima2023drivelm}, and more~\cite{fu2024mobile, driess2023palme, rt22023arxiv}.
Recent vision-language-action (VLA) models~\cite{kim24openvla, pertsch2025fast, goyal2025vla0, kim2025fine, guo2024prediction, chi2023diffusion, li2025unified, du2023learning} have substantially improved a robot’s ability to understand natural language and generate task-level actions under dynamic or unseen environments.

However, an EAI system is more than a VLA model.
In practice, an EAI system consists of a hierarchical control stack with two tightly coupled control loops~\cite{merel2018hierarchical} (\Sect{sec:bg:pipeline}).
At the higher level, a VLA model periodically consumes an instruction and the latest sensory observation to predict an action horizon, i.e., a short sequence of actions.
At the lower level, a model predictive control (MPC) algorithm continuously converts these high-level actions into fine-grained actuator signals.
Moreover, the lower-level controller needs to incorporate recent states, motion constraints, and obstacle information to maintain a safe and smooth trajectory.

These two control levels operate at different frequencies.
VLA inference is compute-intensive but is typically invoked at a relatively low frequency, e.g., 20–60~Hz~\cite{black2026pi, li2025unified, kim24openvla}. 
In contrast, the low-level controller must respond rapidly to changes in the robot state and its environment. 
Depending on the platform and actuator, the control frequency can reach 500 Hz~\cite{herron2024joint, gorbani2025unified, neuman2021robomorphic}.
Meeting low-level control deadlines is often the primary runtime requirement, since deadline misses can introduce trajectory jitter or unsafe behaviors.

\para{Issue.}
Conceptually, these two loops can execute concurrently.
After the VLA model predicts an action horizon, the controller can continuously optimize and execute the current trajectory while the VLA model processes a new observation.
In practice, however, existing implementations have to execute the two workloads \textit{sequentially} due to resource constraints, 
since these mobile platforms, e.g., Nvidia Jetson Orin~\cite{orinsoc} and Thor~\cite{thorsoc}, often have less compute capacity than workstation GPUs.
\Sect{sec:eval:perf} shows that sequential execution can lead to severe quality-of-service (QoS) violations.

\para{Opportunity.}
Although VLA and MPC operate at different frequencies, their dependencies do not require the two control loops to remain serialized.
Once a VLA action becomes available, MPC can continuously generate control commands using the latest robot state while subsequent VLA inference can proceed.
This naturally creates opportunities to overlap the two workloads.

Moreover, VLA and MPC require different GPU resources.
VLA inference is dominated by dense tensor operations that primarily utilize Tensor cores, whereas MPC consists mainly of small matrix and vector operations executed on CUDA cores (\Sect{sec:motiv}).
Their complementary execution characteristics create opportunities for fine-grained co-location on the same GPU.

\para{Challenge.}
However, efficiently co-locating VLA and MPC is a scheduling problem that must account for both control dependencies and GPU resource constraints.
First, MPC jobs have strict deadlines and depend on the VLA-generated actions.
Their execution must be scheduled within the available overlap windows.
Second, different VLA operators exhibit distinct resource demands.
Thus, co-location performance depends heavily on the execution mechanism and GPU resource allocation.
All these factors together determine the overall VLA-MPC execution performance.

\para{Framework.}
To address these challenges, we introduce \proj, a framework for jointly orchestrating VLA inference and MPC computation on a single GPU (\Sect{sec:framework}).
Given a VLA model, an MPC algorithm, a target MPC frequency, and a target platform, \proj first identifies the execution windows in which MPC jobs can overlap with VLA stages based on their data dependencies.
It then constructs hardware-feasible co-location strategies for different VLA operators. 
Lastly, \proj jointly determines the co-location mechanism and MPC workload assignment for each VLA stage, minimizing VLA inference latency while satisfying MPC QoS.

\para{VLA-MPC Co-Location.}
\proj exposes different co-location mechanisms to different VLA operators based on their resource demands.
For linear operations, we observe that linear operations account for most of the VLA execution time and primarily use Tensor cores, while MPC kernels mainly use CUDA cores.
\proj considers three mechanisms where supported by the target GPU: spatial partitioning, stream-based block-level co-location, and warp-specialized intra-block fusion (\Sect{sec:framework:mechanism}).
For a linear-MPC operation pair, the best co-location mechanism depends on both the GPU architecture and workload characteristics.
For attention and non-linear operations, \proj restricts co-location to spatial partitioning to reduce contention for SM-local resources.
To let the optimizer vary the amount of MPC work paired with each VLA operation, \proj groups consecutive QP iterations and partitions MPPI samples into independent path groups.


\para{Co-Location Optimization.}
Given various co-location mechanisms, \proj proposes an optimizer in \Sect{sec:solver} that formulates the VLA-MPC co-location mapping as a constrained optimization problem.
For each candidate mechanism, \proj first uses offline-calibrated performance models to estimate the co-location latency of the VLA stage and its assigned MPC (\Sect{sec:solver:model}).
Using these estimates, \proj selects the co-location mechanism and GPU resource allocation for each VLA stage and determines the number of QP iterations or MPPI paths to assign to it (\Sect{sec:solver:formulation}). 
By solving the constrained optimization, \proj automatically minimizes VLA inference latency subject to workload dependencies, GPU resource constraints, and MPC QoS requirements.

\para{Results.}
We evaluate \proj on two robotic platforms, Nvidia Jetson Orin~\cite{orinsoc} and Thor~\cite{thorsoc}, across three widely adopted VLA inference paradigms~\cite{kim24openvla, kim2025fine, li2025unified}.
Compared with the sequential baseline, \proj improves end-to-end EAI performance by up to 1.4$\times$ and reduces QoS violations by up to 38\%.
Compared with prior co-location strategies~\cite{zhao2022tacker, wu2015enabling}, \proj achieves up to 1.3$\times$ speedup.

In summary, this paper makes the following contributions:
\begin{itemize}
    \item We characterize end-to-end EAI execution on mobile GPUs and expose the potential co-location opportunities between VLA inference and high-frequency MPC computation.
    \item We propose an operator-aware co-location framework that exploits complementary GPU resources through spatial partitioning, block-level co-location, and intra-block fusion.
    \item We formulate VLA–MPC orchestration as a constrained optimization and develop an efficient scheduling solver.
    \item \proj achieves up to 1.4$\times$ end-to-end speedup and 38\% fewer QoS violations over the GPU baselines.
\end{itemize}

\section{Background}
\label{sec:bg}

This section provides the necessary background for EAI systems.
We first introduce the control pipeline of an EAI system with two tightly coupled control loops
(\Sect{sec:bg:pipeline}).
We then explain representative VLA inference paradigms in \Sect{sec:bg:vla}.
Lastly, we introduce two widely used MPC solvers, QP- and MPPI-based MPC, in \Sect{sec:bg:mpc}.

\subsection{Embodied AI Systems}
\label{sec:bg:pipeline}

\begin{figure}[t]
    \centering
    \includegraphics[width=\columnwidth]{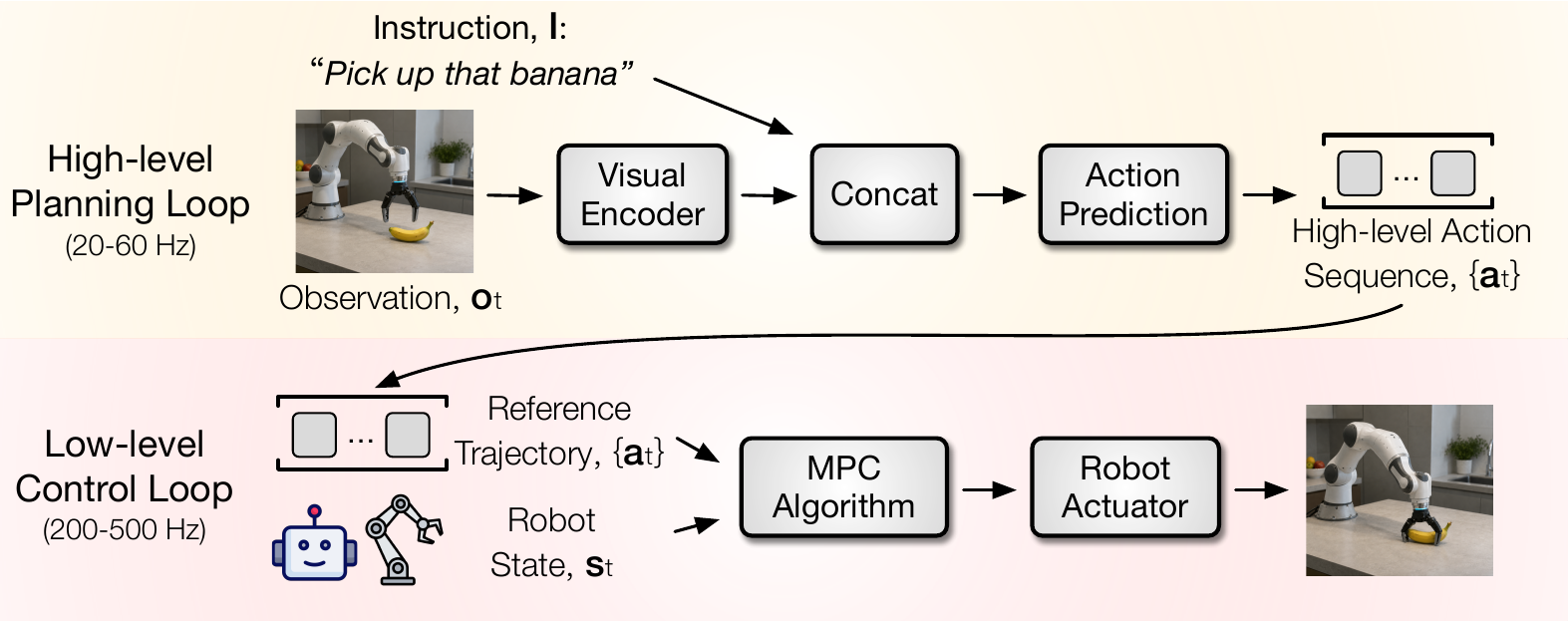}
    \caption{The control loops in EAI systems. The high-level planning loop processes the instruction $\mathbf{I}$ and current observation $\mathbf{o}_t$ to predict a sequence of actions, $\{\mathbf{a}_{t}\}$, at 20-60~Hz. The low-level control loop combines the predicted actions with the current robot state, $\mathbf{s}_t$, and executes MPC algorithms at 200-500~Hz to generate smooth actuator commands.}
    \label{fig:system_workflow}
\end{figure}

As shown in \Fig{fig:system_workflow}, an EAI system typically consists of two tightly coupled control loops: a high-level planning loop and a low-level control loop.
The high-level loop is responsible for understanding task instructions and predicting future actions in the current environment.
It commonly executes a VLA model, which takes a natural language instruction, $\mathbf{I}$, together with the latest sensory observations, $\mathbf{o}_t$, as input and predicts a short horizon, i.e., a sequence of high-level actions, $\{\mathbf{a}_{t}\}$.
Within each iteration, the VLA model includes new sensory observations to generate updated actions.

The goal of the low-level control loop is to translate the action sequence $\{\mathbf{a}_{t}\}$ predicted by the VLA model into executable actuator commands.
Unlike the high-level loop, which primarily focuses on short-horizon planning, the low-level loop must account for robot dynamics, actuator constraints, and nearby obstacles.
Its objective is to ensure that the planned actions can be executed safely and smoothly without exceeding the motor limits or destabilizing the system, e.g., causing the robot to fall.
Therefore, the low-level MPC algorithm continuously refines the high-level action sequence and generates control signals that can be directly issued to the actuators.

Both control loops operate iteratively but at substantially different frequencies.
E.g., the high-level loop typically runs at 20–60~Hz~\cite{black2026pi, li2025unified}, whereas the low-level loop executes at a much higher frequency, 200-500~Hz~\cite{herron2024joint, gorbani2025unified}.
This hierarchical organization allows the robot to combine the semantic reasoning of VLA models with the responsiveness and safety guarantees of conventional MPC algorithms.

\subsection{VLA Models}
\label{sec:bg:vla}

\begin{figure}[t]
    \centering
    \includegraphics[width=\columnwidth]{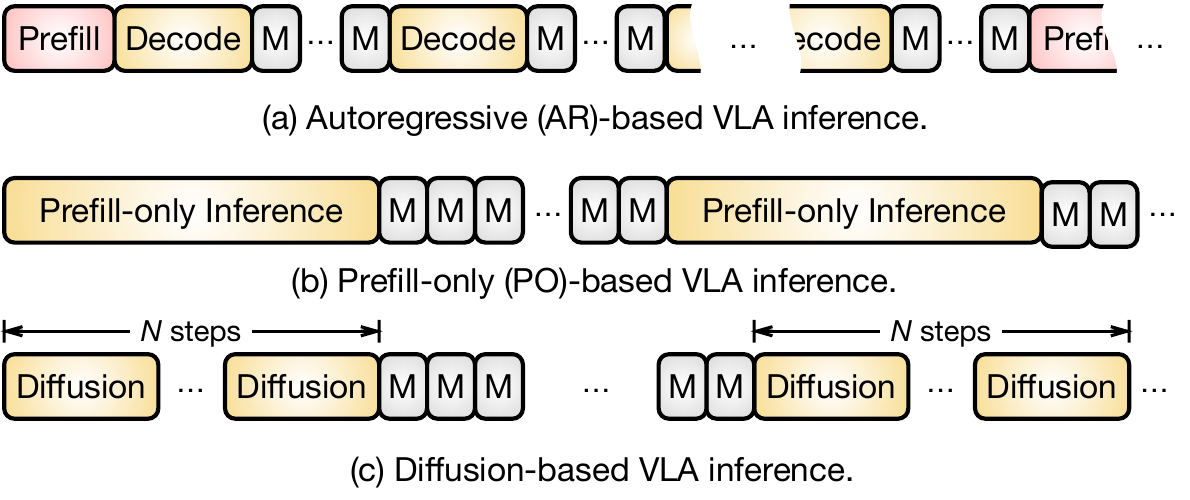}
    \caption{Timelines of representative VLA inference paradigms and their subsequent MPC executions, denoted by \textit{M}. In existing EAI systems, VLA and MPC are executed sequentially.}
    \label{fig:vla_types}
\end{figure}

VLA models serve as the high-level policies to predict short-horizon future actions in an EAI system.
However, existing VLA models~\cite{ma2026survey, zhong2025survey} have different action generation paradigms.
Each offers trade-offs between efficiency and task accuracy. 
We classify their inference processes into three representative paradigms: \textit{autoregressive-based}~\cite{kim24openvla, pertsch2025fast, goyal2025vla0}, \textit{prefill-only-based}~\cite{kim2025fine}, and \textit{diffusion-based}~\cite{guo2024prediction, li2025unified, hu2024video} inference.
Despite these differences, existing baseline EAI systems generally execute VLA inference and MPC computation, denoted by \textit{M}, sequentially on the GPU, as shown in \Fig{fig:vla_types}.

\para{Autoregressive VLA.}
\Fig{fig:vla_types}a shows the pipeline of autoregressive (AR)-based VLA inference.
Each inference starts with a \textit{prefill} stage that encodes the instruction and visual observations.
The model then enters a \textit{decoding} stage.
In this stage, the VLA model generates action tokens one at a time.
Because each decoding step depends on previously generated tokens, the decoding process must execute sequentially.
One action is often represented by multiple tokens, e.g., an arm action is encoded by seven tokens~\cite{kim24openvla}.
Different VLAs could have different action encoding schemes~\cite{yu2025survey}.
Once a complete action is generated, the sequential baseline pauses VLA decoding and invokes MPC to refine the action into low-level controls.
Decoding resumes after the scheduled MPC work completes.

\para{Prefill-Only VLA.}
\Fig{fig:vla_types}b shows the pipeline of prefill-only (PO)-based VLA inference.
A PO-based VLA model predicts a short action chunk in a single prefill stage without a subsequent autoregressive decoding stage.
Consequently, PO-based inference is dominated by highly parallel tensor kernels, e.g., linear operations.
Once an action chunk has been generated, the MPC controller processes it to produce low-level controls before the next VLA inference.

\para{Diffusion-Based VLA.}
\Fig{fig:vla_types}c shows the pipeline of diffusion-based VLA inference.
A diffusion-based VLA model starts from a noisy action sequence and iteratively refines it through $N$ denoising steps.
During each step, the model generates an updated action prediction conditioned on the instruction and visual observations.
As shown in \Fig{fig:vla_types}c, all actions within one denoising step can be processed in parallel.
However, consecutive denoising steps must execute sequentially because each step depends on the output of the previous one.
Once the denoising process completes, the result action chunk is then passed to the MPC controller.
Similar to PO-based VLAs, existing systems execute MPC after VLA inference.

\para{Sequential VLA-MPC Execution.}
Across all three inference paradigms, existing EAI systems treat VLA and MPC as separate GPU workloads and execute them sequentially.
This execution model essentially serializes two control loops on a single resource-constrained mobile GPU.
As we show in \Sect{sec:framework}, this serialization can leave significant GPU resources underutilized, which motivates us to overlap VLA with MPC to reduce end-to-end EAI latency.

\subsection{MPC Algorithms}
\label{sec:bg:mpc}

\begin{figure}[t]
    \centering
    \includegraphics[width=0.95\columnwidth]{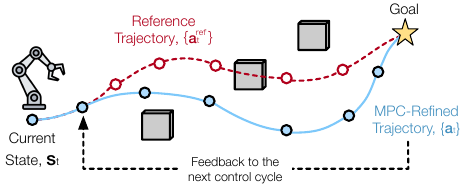}
    \caption{Role of MPC in refining a VLA-generated reference trajectory. Given the current robot state, $\mathbf{s}_{t}$, MPC predicts a trajectory by considering the robotic and environmental constraints. Once the trajectory has been predicted, only the first optimized control input is executed, and the new state, $\mathbf{s}_{t+1}$, is fed back at the next control cycle.}
    \label{fig:mpc_algo}
\end{figure}

As discussed in \Sect{sec:bg:pipeline}, MPC serves as the low-level controller in an EAI system.
As shown in \Fig{fig:mpc_algo}, given the current robot state, $\mathbf{s}_{t}$, and a reference action sequence, $\{\mathbf{a}_{t}\}$, from a VLA model, MPC predicts the robot motion over a short horizon and searches for a sequence of control inputs that can best track the reference trajectory.
Unlike the VLA model, which generates high-level actions, MPC must account for robot dynamics, motor limits, collision constraints, and more~\cite{brito2019model}.
Once the MPC generates a sequence of control inputs, only the first control input of the optimized sequence is issued to the actuators.
At the next control cycle, the MPC then incorporates the latest robot state and repeats the optimization.
This enables the controller to respond rapidly to changes in the robot state and its environment until the target destination is reached.

Existing MPC algorithms employ different optimization solvers.
In general, these solvers can be broadly classified into two representative paradigms: quadratic-programming (QP)-based MPC~\cite{mayne2000constrained, ding2019real} and model predictive path integral (MPPI)-based MPC~\cite{williams2015model, williams2017model}.
QP-based MPC repeatedly solves a constrained optimization problem, whereas MPPI-based MPC searches for a control sequence by sampling and evaluating candidate trajectories.
The following paragraphs explain their basic workflows and characteristics.

\para{QP-Based MPC.}
This type of MPC solver reformulates low-level control as a constrained optimization over a finite prediction horizon.
Given the initial robot state $\mathbf{s}_0$ and a reference trajectory $\{\mathbf{a}_t^\mathrm{ref}\}$, the controller solves,
\begin{equation}
\begin{aligned}
\min_{\{\mathbf{s}\},\,\{\mathbf{u}\}}
\quad &
\sum_{t=0}^{H-1}
\left(
\left\|\mathbf{s}_t-\mathbf{s}^{\mathrm{ref}}_t\right\|_Q^2
+
\left\|\mathbf{u}_t-\mathbf{u}^{\mathrm{ref}}_t\right\|_R^2
\right)
+
\left\|\mathbf{s}_H-\mathbf{s}^{\mathrm{ref}}_H\right\|_{Q_H}^2,
\\
\mathrm{s.t.}\quad &
\mathbf{s}_{t+1}=\mathbf{A}_t\mathbf{s}_t+\mathbf{B}_t\mathbf{u}_t, \quad \mathbf{s}_t\in\mathcal{S},\quad \mathbf{u}_t\in\mathcal{U}.
\end{aligned}
\label{eq:qp_mpc}
\end{equation}
where $H$ is the length of the prediction horizon; $\mathbf{s}_t$ and $\mathbf{u}_t$ are the predicted state and control input; $\mathbf{s}^{\mathrm{ref}}_t$ and $\mathbf{u}^{\mathrm{ref}}_t$ are the reference state and input derived from the reference actions $\{\mathbf{a}_t^\mathrm{ref}\}$.
$\mathbf{A}_t$ and $\mathbf{B}_t$ describe the linearized robot dynamics at timestep $t$.
$Q$, $R$, and $Q_H$ are three matrices to weight the trajectory-tracking, control-effort, and terminal-state costs, respectively.
Lastly, $\mathcal{S}$ and $\mathcal{U}$ describe robot and safety constraints.

QP solvers refine the control sequence through iterative optimization, which is inherently sequential.
Within each iteration, however, operations, e.g., robot-state propagation, residual computation, and variable updates, expose data-level parallelism.
For instance, OSQP~\cite{stellato2018osqp} accelerates convex QPs by exploiting SIMD vector instructions.
Its GPU counterpart, cuOSQP~\cite{schubiger2020gpu}, maps the dominant sparse matrix and vector operations onto CUDA cores to achieve higher parallelism.

\para{MPPI-Based MPC.}
This class of MPC algorithms does not directly solve a constrained optimization problem.
Instead, MPPI approximates the solution by sampling and evaluating a large number of candidate control sequences.
Given the current robot state, $\mathbf{s}_0$, and a reference control sequence, $\{\mathbf{u}_t^\text{ref}\}$, derived from the VLA-generated actions, $\{\mathbf{a}_t^\mathrm{ref}\}$, MPPI samples $K$ perturbed sequences,
\begin{equation}
\mathbf{u}^{(k)}_t = \mathbf{u}_t^\text{ref} + \boldsymbol{\epsilon}^{(k)}_t,
\quad k = 1, \ldots, K,
\label{eq:mppi_sample}
\end{equation}
where $\boldsymbol{\epsilon}^{(k)}_t$ denotes the sampled noise.
Each sampled sequence is then propagated via the robot dynamics model $f(\cdot)$,
\begin{equation}
\mathbf{s}^{(k)}_{t+1} = f( \mathbf{s}^{(k)}_t, \mathbf{u}^{(k)}_t ),
\label{eq:mppi_dynamics}
\end{equation}
to predict a candidate trajectory over a horizon of length $H$.
The cost of each candidate trajectory is evaluated as,
\begin{equation}
L^{(k)} = \sum_{t=0}^{H-1} (
\| \mathbf{s}^{(k)}_t-\mathbf{s}^{\mathrm{ref}}_t \|_Q^2
+ \| \mathbf{u}^{(k)}_t-\mathbf{u}^{\mathrm{ref}}_t \|_R^2 )
+ \| \mathbf{s}^{(k)}_H-\mathbf{s}^{\mathrm{ref}}_H \|_{Q_H}^2,
\label{eq:mppi_cost}
\end{equation}
where $\mathbf{s}^{(k)}_t$ and $\mathbf{u}^{(k)}_t$ are the predicted state and control of the $k$th sampled trajectory.
$L^{(k)}$ is similar to the loss in \Eqn{eq:qp_mpc}.

After evaluating the candidate trajectories, MPPI assigns each trajectory a weight, $w_k$, based on $L^{(k)}$.
The final control sequence is obtained using the weighted sampled noises,
\begin{equation}
\mathbf{u}^{*}_t = \mathbf{u}^{\mathrm{ref}}_t + \sum_{k=1}^{K} w_k \boldsymbol{\epsilon}^{(k)}_t.
\end{equation}
Only the first control input, $\mathbf{u}^{*}_0$, is issued to the actuator, and the process repeats at the next control cycle.

Although the computation within each trajectory is sequential across timesteps, different sampled trajectories are mutually independent and can be evaluated in parallel.
Existing GPU implementations~\cite{williams2015model,williams2017model} therefore map trajectories to different CUDA threads or thread blocks and use parallel reductions to aggregate their costs and update the control sequence.
Thus, MPPI-based MPC exposes path-level parallelism and is well suited to GPU execution.

\section{Motivation}
\label{sec:motiv}

In this section, we explain the need to jointly optimize VLA inference and MPC computation in an end-to-end EAI system.
We first show that both workloads take non-trivial time in the overall execution.
We then characterize their GPU execution behaviors and show the opportunity for concurrent execution.

\begin{figure}[t]
    \centering
    \includegraphics[width=\columnwidth]{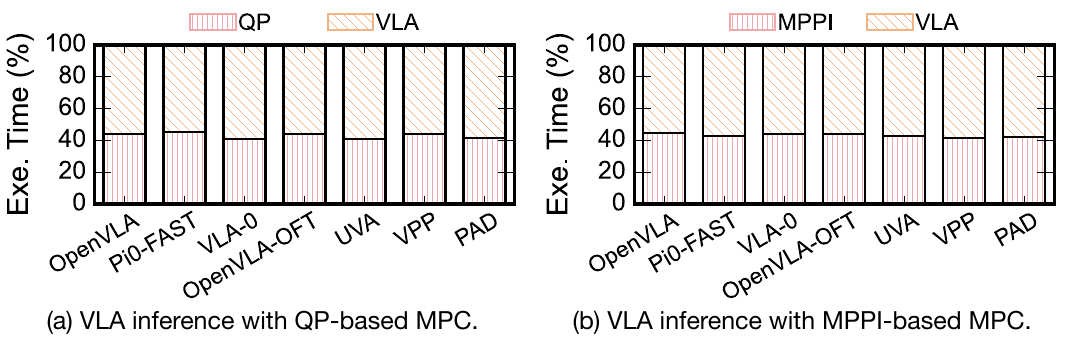}
    \caption{The execution breakdown of seven mainstream  VLA models~\cite{kim24openvla, pertsch2025fast, goyal2025vla0, kim2025fine, guo2024prediction, li2025unified, hu2024video} with two MPCs~\cite{ding2019real, williams2017model} on Orin SoC~\cite{orinsoc}. (a) QP-based MPCs; (b) MPPI-based MPCs.}
    \label{fig:exec_time}
\end{figure}

\para{Execution Breakdown.}
\Fig{fig:exec_time} shows the execution breakdown of seven representative end-to-end EAI workloads across different VLA inference paradigms and MPC solvers.
We measure their latency on a mobile SoC, Nvidia Orin~\cite{orinsoc}, where VLA inference and low-level MPC computation are executed sequentially.
Both VLA inference and MPC computations take non-trivial execution time.
For instance, with the QP-based MPC, VLA inference and MPC account for 57\% and 43\% of the end-to-end latency, respectively.
Although each MPC execution is relatively lightweight, its high frequency makes its cumulative execution non-trivial.

The result shows that accelerating only VLAs leaves a substantial fraction of the end-to-end runtime untouched.
Moreover, simply interleaving the two workloads can delay latency-critical MPC execution and lead to severe QoS violations.
Thus, improving end-to-end performance requires optimizing both VLA and MPC.

\begin{figure}[t]
    \centering
    \includegraphics[width=\columnwidth]{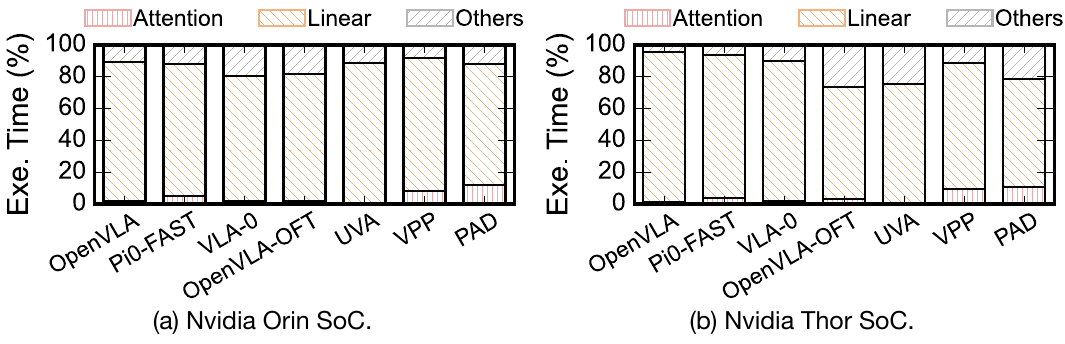}
    \caption{The operator-level breakdown of VLA models~\cite{kim24openvla, pertsch2025fast, goyal2025vla0, kim2025fine, guo2024prediction, li2025unified, hu2024video} on two Nvidia mobile SoCs, Orin~\cite{orinsoc} and Thor~\cite{thorsoc}. Linear operations are the main bottlenecks.}
    \label{fig:gpu_utils}
\end{figure}

\para{Workload Characteristics.}
Although VLA inference and MPC computation both execute on the GPU, they show different computational characteristics.
VLA is dominated by dense matrix multiplications, which are typically mapped to Tensor cores rather than general-purpose CUDA cores.
\Fig{fig:gpu_utils} shows the breakdown of different operators in VLA models.
For instance, on Orin SoC~\cite{orinsoc}, linear layers in VLA models account for 80\% of the VLA execution time and primarily consist of compute-intensive matrix multiplications.

\begin{figure}[t]
    \centering
    \includegraphics[width=\columnwidth]{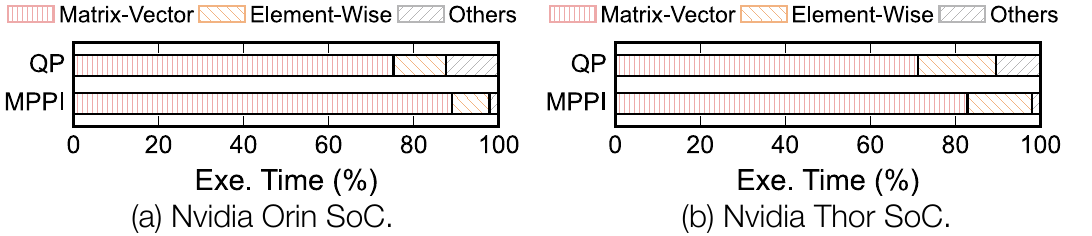}
    \caption{The operator breakdown of two MPC algorithms~\cite{ding2019real, williams2017model} on two Nvidia mobile SoCs, Orin~\cite{orinsoc} and Thor~\cite{thorsoc}.}
    \label{fig:mpc_utils}
\end{figure}

In contrast, MPC algorithms are predominantly executed on CUDA cores.
For instance, QP-based MPC mainly consists of sparse matrix-vector multiplications, residual computations, and iterative variable updates.
MPPI-based MPC propagates robot states and evaluates costs across a large number of sampled trajectories.
These operations often involve small tensor shapes and element-wise computation.
\Fig{fig:mpc_utils} shows the execution time breakdown of MPC algorithms on two SoCs.
On average, matrix-vector and element-wise operations take 96\% of the total MPC execution time.
Thus, MPC workloads are more suitable for CUDA cores.

Naturally, the execution characteristics of these two workloads create a co-location opportunity.
While a VLA kernel occupies the Tensor cores, MPC computations can use otherwise idle CUDA cores.
Overlapping the two workloads can potentially increase GPU utilization and reduce end-to-end latency without additional hardware resources.
One caveat is that Tensor cores and CUDA cores are not fully independent: they still share streaming multiprocessor (SM) resources, e.g., registers and shared memory.
Effectively orchestrating their execution is key to the overall performance.

\para{Summary.}
End-to-end EAI systems exhibit two key characteristics.
First, both the VLA loop and the high-frequency MPC loop contribute non-trivially to the overall execution latency.
Second, the two workloads often rely on different GPU execution resources: VLAs primarily use Tensor cores, whereas MPCs mainly use CUDA cores.
Together, these characteristics motivate fine-grained concurrent execution to improve end-to-end EAI.
\section{\proj Framework}
\label{sec:framework}

The goal of \proj is to improve the overall VLA performance while preserving the QoS of MPC controls.
To achieve this goal, this section first describes the possible mechanisms that overlap these two workloads (\Sect{sec:framework:mechanism}).
Then, we describe the eligible co-location mechanisms for different VLA operators (\Sect{sec:framework:workloads}). 

\subsection{Overview}
\label{sec:framework:ov}

\begin{figure}[t]
    \centering
    \includegraphics[width=\columnwidth]{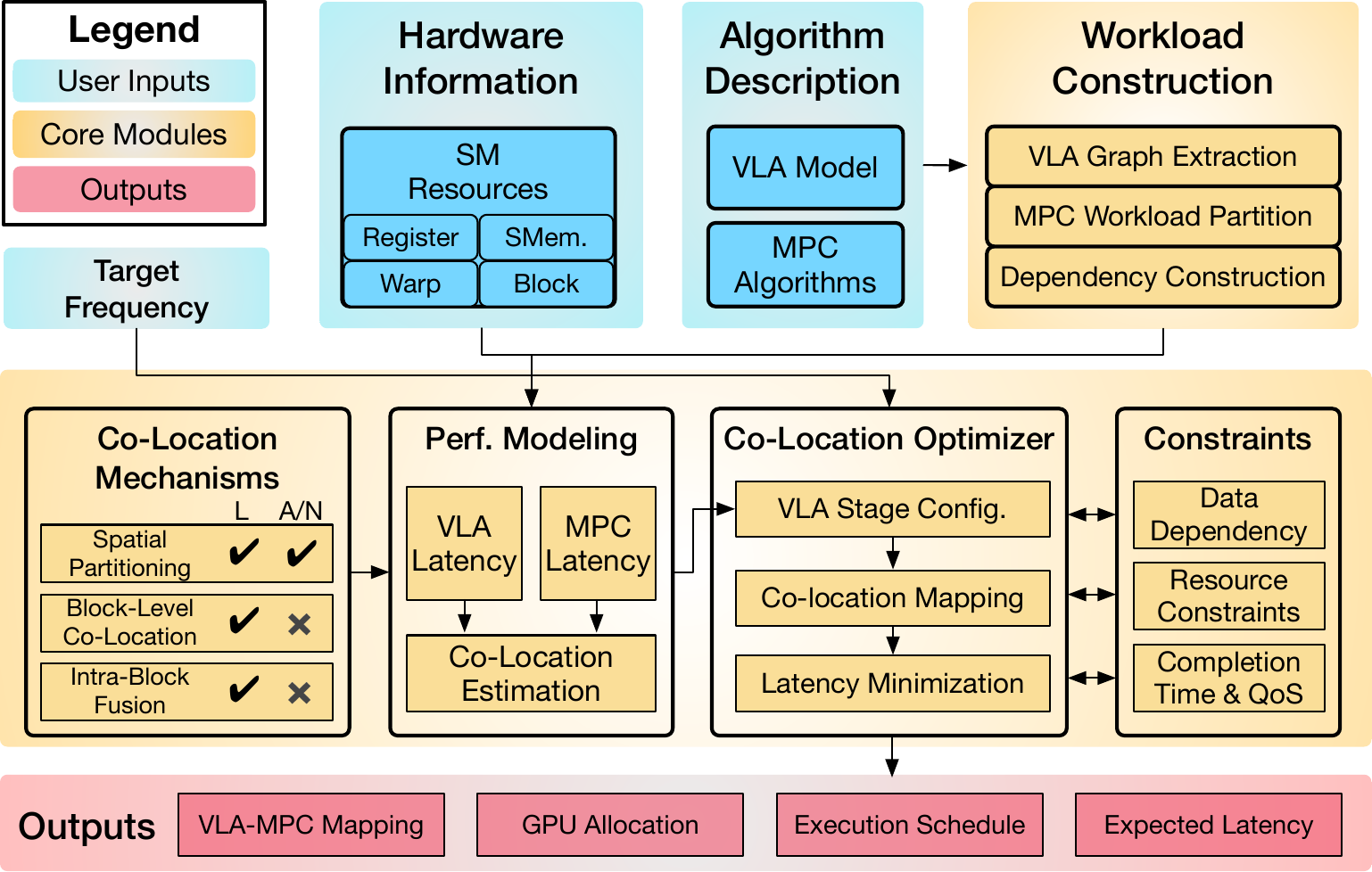}
    \caption{Overview of the \proj framework. \proj constructs VLA stages and MPC work groups, then enumerates hardware-feasible co-location configurations by VLA operator class. The performance model estimates VLA-stage and MPC-chunk completion times for each candidate. The optimizer uses these estimates to select stage configurations and map MPC work to VLA stages while respecting dependencies, GPU resource reservations, and MPC deadlines.}
    \label{fig:ov_framework}
\end{figure}

\Fig{fig:ov_framework} gives an overview of \proj.
Given a VLA and an MPC algorithm, a target control frequency, and a GPU platform, \proj automatically derives a VLA-MPC co-location schedule that improves VLA efficiency while satisfying the QoS of the MPC controller.

\proj consists of three major components.
First, \proj identifies the execution windows in which high-frequency MPC can overlap with subsequent VLA inference, as described in \Sect{sec:framework:opportunity}.
These opportunities depend on the inference paradigm of the VLA model and the dependencies between the two control loops.

Second, \proj constructs candidate co-location configurations based on the resource characteristics of each VLA operator.
As shown in \Sect{sec:framework:mechanism} and \Sect{sec:framework:workloads}, \proj considers three co-location mechanisms: spatial partitioning, block-level co-location, and intra-block fusion.
Linear operations can use all three mechanisms, while attention and non-linear operations use spatial partitioning to avoid contention with latency-critical MPC computation.
Meanwhile, QP and MPPI workloads are partitioned into schedulable chunks, i.e., iteration groups and sampled-path groups, respectively.

Finally, \proj proposes the co-location optimizer in \Sect{sec:solver} to determine the execution configuration of each VLA stage and the co-location placement of MPC computation across the VLA timeline.
The optimizer then uses offline performance models to estimate candidate execution latencies and minimizes the end-to-end VLA inference latency while enforcing constraints, e.g., MPC deadlines, workload dependencies, and GPU resources.
The final output specifies the co-location mechanism, resource configuration, and kernel mapping between MPC and VLA.

\begin{figure}[t]
    \centering
    \includegraphics[width=\columnwidth]{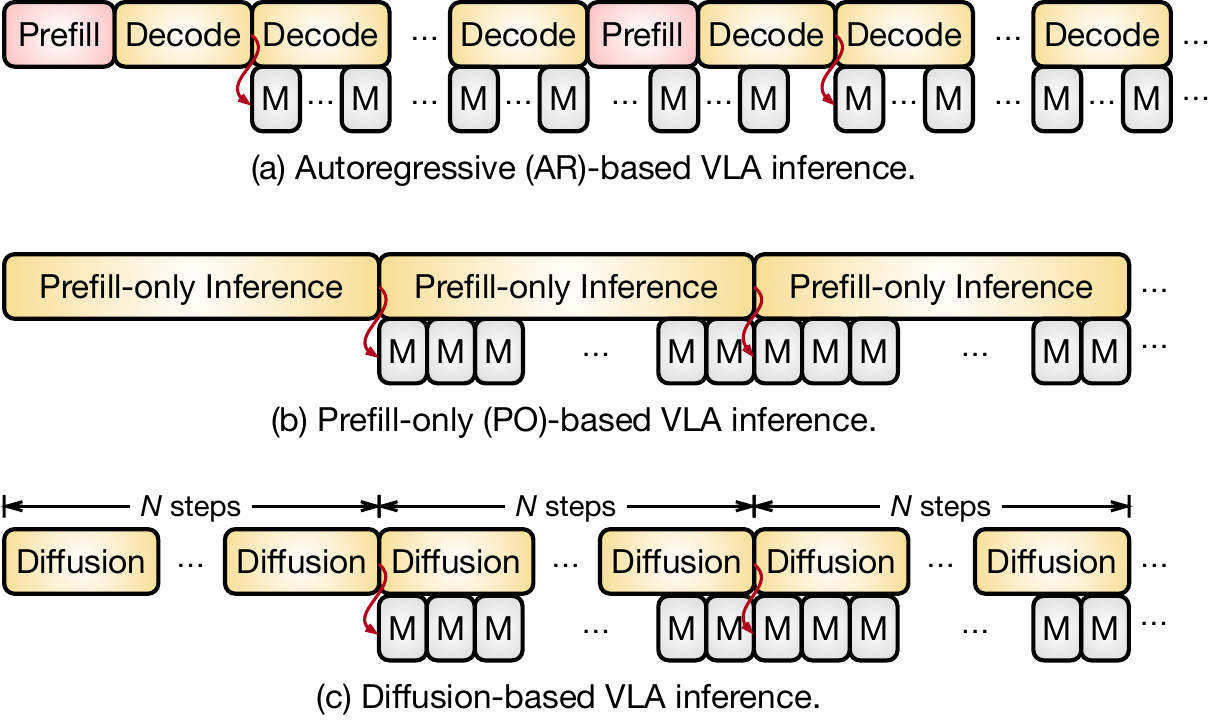}
    \caption{Co-location opportunities between VLA inference and MPC computation across three representative VLA paradigms. Once the required VLA action becomes available, MPC computation, denoted by \textit{M}, can overlap with subsequent VLA execution. Red arrows indicate dependencies between the two control levels.}
    \label{fig:colocate_oppo}
\end{figure}

\subsection{VLA-MPC Co-Location Opportunities}
\label{sec:framework:opportunity}

\para{Opportunities.}
Although existing EAI systems execute VLA and MPC sequentially, their dependencies do not require these two control loops to remain serialized throughout execution.
Once a VLA model produces the action required by the low-level controller, the corresponding MPC computation can proceed while subsequent VLA inference can continue.
Each VLA invocation uses the latest available observation, while MPC continuously generates control commands using the current robot state and the latest available reference trajectory.
Once a new VLA action becomes available, the reference is updated for subsequent MPC invocations.
This creates an opportunity to overlap the two control loops, as shown in \Fig{fig:colocate_oppo}.

\Fig{fig:colocate_oppo} shows these overlap opportunities for the three representative VLA inference paradigms introduced in \Sect{sec:bg:vla}.
For AR-based VLA models, once a complete action becomes available during decoding, its MPC computation can overlap with subsequent decoding steps.
For PO-based VLA models, MPC computation associated with the current action chunk can overlap with the next VLA inference.
Similarly, for diffusion-based VLA models, once the denoising process produces an action chunk, its MPC computation can overlap with subsequent diffusion inference.
Although the available overlap windows differ across inference paradigms, all three expose opportunities to execute VLA and MPC concurrently.

These overlap windows determine when MPC can execute concurrently with VLA.
However, the optimal way to achieve this execution overlap depends on the resource characteristics of both the VLA operator and the MPC within each window.
We next introduce the mechanisms exploited by \proj in \Sect{sec:framework:mechanism}.

\subsection{Co-Location Mechanisms}
\label{sec:framework:mechanism}

\begin{figure}[t]
    \centering
    \includegraphics[width=\columnwidth]{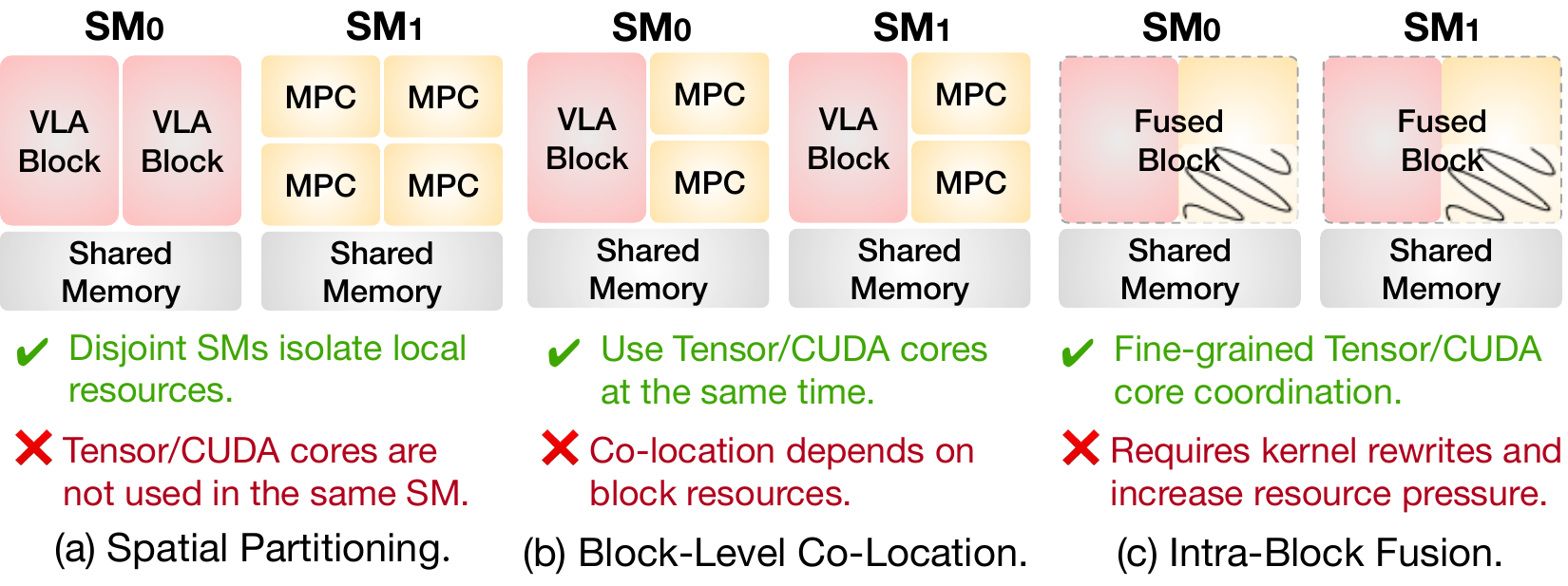}
    \caption{GPU resource-sharing trade-offs for three VLA–MPC co-location mechanisms.}
    \label{fig:colocation}
\end{figure}

Given an overlap window between VLA inference and MPC computation, the next question is how these two workloads should be co-located.
Here, \proj considers three co-location mechanisms with different granularities: \textit{spatial partitioning}, \textit{block-level co-location}, and \textit{intra-block fusion}.
\Fig{fig:colocation} shows their examples.

\para{Spatial Partitioning.}
One approach to co-executing two workloads is to use existing GPU spatial partitioning tools, e.g., MPS~\cite{mps_doc}, to divide SMs between VLA and MPC.
As shown in \Fig{fig:colocation}a, one SM group executes VLA kernels while another executes MPC kernels.
Because the two workloads are isolated at the SM level, spatial partitioning avoids direct contention for SM-local resources and requires minimal kernel modification.
However, it cannot exploit complementary compute resources within the same SM.
E.g., SMs assigned to VLA may leave part of their CUDA cores underutilized, while SMs assigned to MPC may leave Tensor cores idle.

\para{Block-Level Co-Location.}
This mechanism keeps VLA and MPC as separate GPU kernels while allowing thread blocks from the two kernels to co-locate on the same SM.
As shown in \Fig{fig:colocation}b, each block preserves its original thread organization, register allocation, shared-memory allocation, and synchronization domain.
Separate CUDA streams allow the GPU to launch blocks from another kernel once all blocks of the first kernel have been issued; whenever an SM has sufficient resources, blocks from the second kernel can be co-located on that SM.
If these two kernels' blocks leverage different compute units, this co-location can achieve the Tensor-CUDA core overlap between VLA and MPC.
Nevertheless, co-located blocks still compete for shared SM resources.
The achievable concurrency depends on their block-level resource footprints.

\para{Intra-Block Fusion.}
To exploit the parallelism between Tensor cores and CUDA cores, another approach is to fuse VLA and MPC computations into a single thread block, a.k.a. warp specialization (WASP)~\cite{crago2024wasp}.
Different warps within the thread block are assigned to different workloads, allowing their execution to be coordinated at a finer granularity.
Prior studies, e.g., Tacker~\cite{zhao2022tacker}, partition the threads of a fused block into two groups: one performs Tensor core computation, while the other performs general-purpose CUDA-core computation.
This mechanism provides fine-grained control over synchronization, resource sharing, and execution order within a block.
However, intra-block fusion must accommodate the execution state and on-chip resource requirements of both workloads, which can increase register and shared-memory pressure.
It also requires additional synchronization and kernel restructuring.
Recent GPU architectures~\cite{blackwell_doc} provide mechanisms, e.g., register reallocation via \texttt{setmaxnreg} and asynchronous execution support, that can reduce some of the overheads for warp-specialized pipelines.

\subsection{VLA-MPC Co-Location Strategy}
\label{sec:framework:workloads}

In \Sect{sec:motiv}, we show that different VLA operators exhibit different GPU resource characteristics.
Therefore, applying a single co-location mechanism throughout VLA execution can either leave GPU resources underutilized or introduce contention that causes MPC to miss its QoS deadline.
Here, \proj adopts different co-location mechanisms for VLA operators.

\para{Linear Operations.}
\Fig{fig:gpu_utils} shows that linear operations take roughly 80\% of VLA execution time and are dominated by dense matrix multiplications using Tensor cores.
In contrast, MPC computation executes on general-purpose CUDA cores.
This makes linear operations suitable for fine-grained VLA-MPC co-location.
Therefore, when MPC overlaps with a linear operation, \proj considers all three mechanisms from \Sect{sec:framework:mechanism}.
The final mechanism and its resource configuration are determined by the co-location optimizer in \Sect{sec:solver} based on the workload characteristics and target GPU.

\para{Other Operations.}
In contrast, attention and non-linear operations are less suitable for fine-grained co-location with MPC on the same SM.
Although attention leverages Tensor-core computation, it also places high pressure on memory bandwidth and shared memory.
On the other hand, non-linear operations primarily execute on general-purpose CUDA cores and thus compete directly with MPC for the same execution resources.
In both cases, sharing an SM with MPC can introduce resource contention.
\proj therefore only uses \textit{spatial partitioning} when MPC overlaps with these operators.

In summary, \proj considers all three mechanisms for linear operations.
For attention and non-linear operations, \proj restricts co-location to spatial partitioning to reduce SM resource contention.
Next, we describe how different MPC algorithms are divided into small execution chunks and co-located with VLA operations.

\para{QP-Based MPC Partitioning.}
As described in \Sect{sec:bg:mpc}, QP-based MPC refines a control sequence through an iterative solver.
Because each iteration depends on the output of the previous one, the iterations must preserve their original execution order.
To make the QP workload predictable for co-location, \proj executes each MPC invocation with a fixed iteration budget.
This budget is sufficient for the QP problems to converge at the scales in our EAI workloads.
\proj then groups consecutive QP iterations into small chunks and co-locates one chunk of iterations with a VLA operation.
The optimizer in \Sect{sec:solver} determines the number of QP iterations assigned to each operation and its co-location mechanism.

\para{MPPI-Based MPC Partitioning.}
Unlike QP-based MPC, MPPI-based MPC evaluates a fixed number of sampled control paths.
Because different paths are mutually independent, they can be evaluated in parallel.
\proj then partitions the sampled paths into groups and co-locates each path group with a VLA operator.
The paths within a group are distributed across GPU threads, allowing their evaluation to overlap with VLA operations.
The co-location optimizer in \Sect{sec:solver} determines how many sampled paths are scheduled within each VLA operator and selects the corresponding co-location mechanism based on the VLA operator and GPU resources.

\section{Co-Location Optimization}
\label{sec:solver}

Our co-location strategy in \Sect{sec:framework} exposes opportunities to overlap VLA with high-frequency MPC kernels.
However, the optimal co-location schedule depends on multiple factors, e.g., the VLA model, MPC algorithm, target control frequency, and GPU platform.
To address this trade-off, \proj formulates VLA-MPC co-location as a constrained optimization problem.
The optimizer minimizes VLA inference latency while enforcing the control deadline of every MPC invocation as a hard QoS constraint.
\Sect{sec:solver:model} first introduces our analytical performance models.
\Sect{sec:solver:formulation} then uses these models to determine the co-location configuration of each VLA stage and the mapping of MPC computation across the VLA timeline.

\subsection{Performance Modeling}
\label{sec:solver:model}

To optimize VLA-MPC co-location, \proj requires latency estimates for its standalone and co-located GPU executions.
We therefore calibrate analytical models offline for each target GPU.

\begin{figure}[t]
\centering
\begin{minipage}[t]{0.48\columnwidth}
  \centering
  \includegraphics[width=\columnwidth]{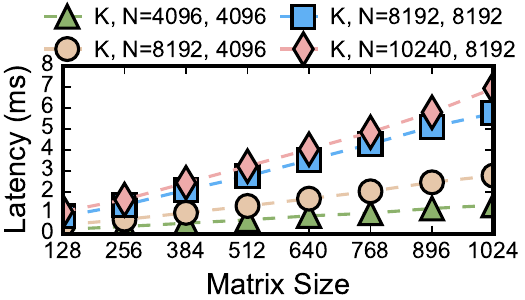}
  \caption{The measured linear layer latency with respect to the input size.}
  \label{fig:mlp_model}
\end{minipage}
\hspace{2pt}
\begin{minipage}[t]{0.48\columnwidth}
  \centering
  \includegraphics[width=\columnwidth]{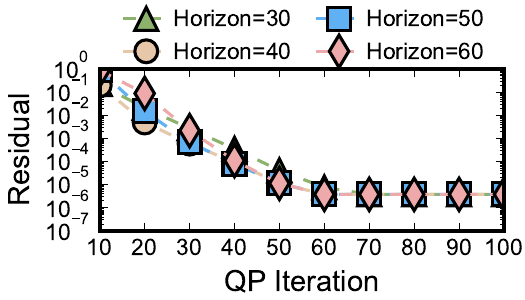}
  \caption{QP loss versus iteration count. QP instances converge in 80 iterations.}
  \label{fig:qp_model}
\end{minipage}
\end{figure}

\para{Standalone Linear Layer Kernels.}
We first describe the standalone latency model for VLA linear layers, which take the majority of VLA execution time and are the primary target for fine-grained co-location.
Other VLA operators, e.g., attention and non-linear operations, can be modeled using the same methodology.
The standalone latency of the $i$th linear layer kernel, $t_{\text{linear}, i}^{V}$, is,
\begin{equation}
    t_{\text{linear}, i}^{V} =
    \cF^{V}(\mathbf{d}_i, \mathbf{c}_i, \mathbf{g}),
\label{eq:vla_model}
\end{equation}
where $\mathbf{d}_i$ is the matrix dimensions, $\mathbf{c}_i$ defines the kernel configuration, e.g., the number of PTB workers, and $\mathbf{g}$ denotes the target GPU platform.
As shown in \Fig{fig:mlp_model}, the kernel latency varies predictably with the matrix dimensions.
Based on these offline profiling results, we can estimate the standalone latency of linear kernels.

\begin{figure}[t]
    \centering
    \includegraphics[width=1\columnwidth]{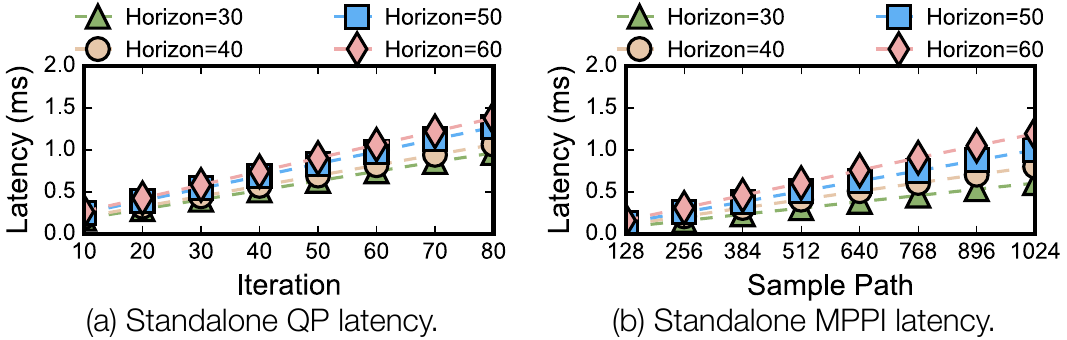}
    \caption{The standalone latency of QP and MPPI with respect to iteration counts, sample paths, and predictive horizon.}
    \label{fig:mpc_latency}
\end{figure}

\para{Standalone QP Kernels.}
QP-based MPC refines its solution through an iterative solver.
To make its workload size predictable, \proj sets a fixed iteration budget for each MPC invocation.
We use a budget of 80 iterations, which was sufficient for convergence on the evaluated QP instances (\Fig{fig:qp_model}).
For performance modeling, we use the number of consecutive QP iterations as the input size of a standalone QP batch (\Fig{fig:mpc_latency}a).
We then model the latency of a QP batch with $n_b^{Q}$ iterations as,
\begin{equation}
 t_\text{MPC,b}^{Q} = \cF^{Q}(\mathbf{p}_b^{Q},\mathbf{g}) \cdot n_b^{Q},
\label{eq:qp_model}
\end{equation}
where $n_b^{Q}$ denotes the number of consecutive iterations in the $b$th batch, and $\mathbf{p}_b^{Q}$ describes the QP problem and kernel configuration.

\para{Standalone MPPI Kernels.}
For MPPI-based MPC, the dominant computation is evaluating sampled control paths over a prediction horizon.
\Fig{fig:mpc_latency}(b) shows that the standalone MPPI latency increases with the number of sampled paths.
Since sampled paths are mutually independent, they can be evaluated concurrently on GPU.
The standalone latency, $t_\text{mpc,b}^{P}$, of a batch of $n_b^{P}$ MPPI paths is,
\begin{equation}
t_\text{mpc,b}^{P} = \cF^{P}(\mathbf{p}_b^{P},\mathbf{g}) \cdot \lceil n_b^{P}/n_g \rceil,
\label{eq:mppi_model}
\end{equation}
where $n_b^{P}$ denotes the number of sampled paths within a batch, $n_g$ is the number of sampled paths that can execute concurrently on GPU $\mathbf{g}$, and $\mathbf{p}_b^{P}$ describes the algorithmic setup and kernel configuration.

\begin{figure}[t]
    \centering
    \includegraphics[width=1\columnwidth]{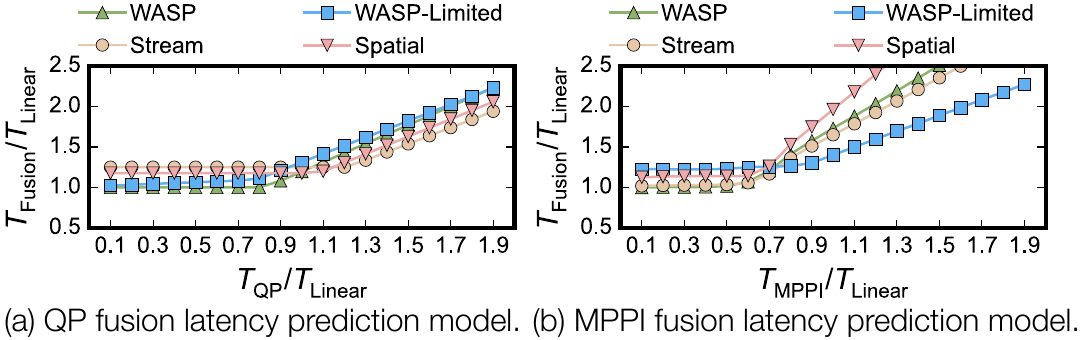}
    \caption{The normalized co-location latency of QP and MPPI with respect to the load ratio.}
    \label{fig:fusion_latency}
\end{figure}
\para{Co-Location Latency.}
Given the standalone estimates above, we model concurrent VLA-MPC execution using the load-ratio formulation from Tacker~\cite{zhao2022tacker}.
For VLA stage $V_i$ and MPC batch $b$ of algorithm type $M\in\{Q,P\}$, we define the ratio,
\begin{equation}
    \rho_{i,b}^{M}
    = {t_{\mathrm{MPC},b}^{M}}/{t_i^{V}}.
\label{eq:fusion_load_ratio}
\end{equation}
Here, $t_i^{V}$ is the standalone latency of the VLA stage, including attention and non-linear stages eligible for spatial partitioning. Both standalone latencies are estimated with each workload running alone on the full target GPU.
We normalize completion latencies by $t_i^{V}$ and fit a piecewise-linear function of $\rho$:
\begin{equation}
    \hat{\eta}^{M,m}(\rho)
    =
    \begin{cases}
        \alpha_{1}^{M,m},
        & \rho\leq\rho_{\mathrm{th}}^{M,m},\\
        \alpha_{2}^{M,m}\rho+\beta_{2}^{M,m},
        & \rho>\rho_{\mathrm{th}}^{M,m}.
    \end{cases}
\label{eq:fusion_ratio_model}
\end{equation}
Here, $m$ denotes the co-location mechanism. We calibrate the coefficients and threshold offline for each resource configuration.
\Fig{fig:fusion_latency} shows the fit for different co-location mechanisms on linear layers.

For intra-block fusion, the VLA and MPC workloads complete together. The estimated latency can be expressed as,
\begin{equation}
    \hat{t}_{i,b}^{M,m}
    =
    t_i^{V}\,\hat{\eta}^{M,m}\!\left(\rho_{i,b}^{M}\right).
\label{eq:fusion_model}
\end{equation}
For spatial partitioning and block-level co-location, we measure VLA-stage and MPC-chunk completion separately from the common launch time and fit one curve for each.
Multiplying each fitted value by $t_i^V$ gives the corresponding candidate latency, $\hat{T}_{k,o}^{V}$ or $\hat{T}_{k,o}^{M}$, for configuration $o$ of stage $V_k$. For intra-block fusion, both latencies equal \Eqn{eq:fusion_model}.
Note that the MPC-chunk latency excludes the final MPPI reduction, which \Sect{sec:solver:formulation} schedules separately.

\subsection{Optimization Formulation}
\label{sec:solver:formulation}

With the performance models in \Sect{sec:solver:model}, we formulate VLA-MPC co-location as a constrained optimization problem.
For clarity, the window covers one VLA inference and its associated MPC jobs.
The formulation can naturally extend to multiple VLA inference steps.

\para{Execution Model.}
We first represent VLA inference as a sequence of schedulable stages,
\begin{equation}
    \mathcal{V}=\{V_0,V_1,\ldots,V_{K-1}\},
\end{equation}
where a stage can correspond to a complete operator, e.g., an attention or non-linear kernel, or a partition of a large linear operation.
For each stage $V_k$, $s_k^V$ and $e_k^V$ denote its start and completion times, respectively.
Let $r^V$ denote the fixed time at which the VLA input is available, and require $s_0^V\geq r^V$. The total VLA latency is,
\begin{equation}
    T_{\mathrm{VLA}} = e_{K-1}^V-r^V.
\label{eq:vla_latency}
\end{equation}


Meanwhile, MPC executes periodically according to the target control frequency.
Let $\mathcal{J}$ denote the MPC jobs within the scheduling window.
Let $\mathcal J_{\mathrm{QP}}$ and $\mathcal J_{\mathrm{MPPI}}$ denote the sets of QP and MPPI jobs, respectively.
We measure all times relative to the release of MPC job $J_0$.
For each job $J_j$, its release time $r_j$ and deadline $d_j$ are,
\begin{equation}
    r_j=j t_P, \qquad d_j=r_j+t_P,
\label{eq:mpc_release_deadline}
\end{equation}
where $t_P = 1/f_{\mathrm{MPC}}$ is the target control period.
Let $A_j$ denote the time when the VLA action required by $J_j$ becomes available.
If stage $V_{a(j)}$ produces the action within the window, $A_j=e_{a(j)}^V$. Otherwise, $A_j$ is known before the window.
Each job requires a fixed amount of computation, $W_j>0$, e.g., a fixed number of QP iterations or MPPI sampled paths.
Let $C_j$ denote the time when the complete result of $J_j$ becomes available to the robot controller.

\para{Variables.}
For each VLA stage $V_k$, \proj constructs offline a set of legal execution configurations, $\mathcal{O}_k$.
Each configuration $o\in\mathcal{O}_k$ specifies a co-location mechanism, GPU resource allocation, the amount of MPC work $w_{k,o}$ executed with the stage, the predicted VLA latency $\hat{T}_{k,o}^{V}$, and the predicted MPC completion latency $\hat{T}_{k,o}^{M}$.
Here, $\hat{T}_{k,o}^{M}$ measures the time from the start of $V_k$ until the assigned MPC chunk completes, excluding the final MPPI reduction.
A standalone VLA configuration has $w_{k,o}=0$ and $\hat{T}_{k,o}^{M}=0$.

Let $y_{k,o}$ and $z_{j,k}$ be binary variables indicating configuration selection and job assignment, respectively. Let $x_{j,k}\in\mathbb Z_{\geq0}$ count the QP iterations or MPPI paths from $J_j$ assigned to $V_k$.
Let $e_{j,k}^{M}$ denote the completion time of this MPC work.
Let $u_j$ denote the reduction start time for an MPPI job.

The complete set of optimization variables is,
\begin{equation*}
    \Theta =
    \bigl\{
        y_{k,o},\,x_{j,k},\,z_{j,k},\,
        s_k^V,\,e_k^V,\,e_{j,k}^M,\,b_k,\,u_j,\,C_j
    \bigr\},
\end{equation*}
where $b_k$ is the end of the GPU reservation for stage $V_k$, $C_j$ is the completion time of job $J_j$, and $u_j$ is defined only for MPPI jobs. The constraints below define the completion and reservation times.

\para{Objective.}
The primary QoS requirement is that every MPC invocation completes before its control deadline.
Subject to this requirement, \proj minimizes VLA inference latency:
\begin{align}
    \min
    \quad
    T_{\mathrm{VLA}}
    &=
    e_{K-1}^V-r^V.
\label{eq:objective}
\end{align}

Next, we describe four categories of constraints:
\textit{configuration and workload constraints},
\textit{data-dependency constraints},
\textit{execution and resource constraints}, and
\textit{MPC QoS constraints}.

\para{Configuration and Workload Constraints.}
Each VLA stage selects exactly one execution configuration:
\begin{equation}
    \sum_{o\in\mathcal{O}_k}y_{k,o}=1,
    \qquad \forall k.
\label{eq:config_selection}
\end{equation}
Its execution latency can be computed as,
\begin{equation}
    T_k^V
    =
    \sum_{o\in\mathcal{O}_k}
    y_{k,o}\hat{T}_{k,o}^{V},
    \qquad
    e_k^V=s_k^V+T_k^V.
\label{eq:vla_stage_latency}
\end{equation}

Each MPC job must complete its entire workload:
\begin{equation}
    \sum_k x_{j,k}=W_j,
    \qquad \forall j.
\label{eq:work_conservation}
\end{equation}
Meanwhile, the total MPC workload assigned to stage $V_k$ must match the workload associated with its selected configuration:
\begin{equation}
    \sum_jx_{j,k}
    =
    \sum_{o\in\mathcal{O}_k}
    w_{k,o}y_{k,o},
    \qquad \forall k.
\label{eq:work_config_coupling}
\end{equation}
We couple workload and assignment with,
\begin{equation}
    z_{j,k} \leq x_{j,k} \leq W_j z_{j,k}, \qquad \forall j,k.
\label{eq:job_stage_coupling}
\end{equation}
Each stage pairs with at most one MPC job:
\begin{equation}
    \sum_j z_{j,k}\leq 1,
    \qquad \forall k.
\label{eq:single_job_stage}
\end{equation}

\para{Data-Dependency Constraints.}
Next, the optimized schedule must preserve the dependencies.
An MPC job may start only after its periodic release and the availability of its required VLA action:
\begin{equation}
    z_{j,k}=1
    \;\Rightarrow\;
    \bigl(s_k^V\geq r_j\ \land\ s_k^V\geq A_j\bigr),
    \qquad \forall j,k.
\label{eq:release_constraint}
\end{equation}
For QP, consecutive iteration groups consume the preceding group's solver state in order.
In contrast, MPPI path groups evaluate independent sets of sampled paths.

\para{Execution and Resource Constraints.}
When MPC work from job $J_j$ is assigned to stage $V_k$, its completion time is determined by the selected co-location configuration.
Each candidate configuration represents a co-launch of the VLA stage and its assigned MPC work at $s_k^V$. For every $j,k$ and $o\in\mathcal O_k$,
\begin{equation}
    z_{j,k}=1,\;y_{k,o}=1
    \;\Rightarrow\;
    e_{j,k}^{M}
    =
    s_k^V+\hat{T}_{k,o}^{M}.
\label{eq:mpc_completion}
\end{equation}
Here, the two workloads can complete separately under spatial partitioning and block-level co-location.
For intra-block fusion, it equals the fused-stage latency.

The GPU is reserved for each selected VLA-MPC pair until both workloads complete.
Let $b_k$ be the end of this reservation. Then,
\begin{equation}
    b_k=s_k^V+
    \sum_{o\in\mathcal O_k} y_{k,o}
    \max\!\left\{\hat T_{k,o}^V,\hat T_{k,o}^M\right\},
    \qquad \forall k.
\end{equation}
The maximum is computed offline for each candidate.
The next stage launches only after the reservation ends:
\begin{equation}
    s_{k+1}^V\geq b_k,
    \qquad k=0,\ldots,K-2.
\label{eq:vla_dependency}
\end{equation}
The VLA output remains available at $e_k^V$, even when its paired MPC finishes later.
Before optimization, candidate construction checks operator eligibility, resource limits, and runtime support.

\para{MPC QoS Constraints.}
We assume that the final QP iteration group makes the control result available. 
The QP job completion time is the latest completion time among its iteration groups:
\begin{equation}
    C_j
    =
    \max_{k:z_{j,k}=1}
    e_{j,k}^{M},
    \qquad \forall j\in\mathcal J_{\mathrm{QP}}.
\label{eq:qp_job_completion}
\end{equation}

For MPPI-based MPC, all sampled-path groups must complete before the final reduction.
Let $T_j^R$ denote the profiled latency of the reduction when it runs alone on the GPU.
For each $j\in\mathcal J_{\mathrm{MPPI}}$,
\begin{equation}
    z_{j,k}=1\;\Rightarrow\;u_j\geq e_{j,k}^{M},
    \quad \forall k,
    \qquad
    C_j=u_j+T_j^R.
\label{eq:mppi_job_completion}
\end{equation}
Each reduction reserves the GPU over $[u_j,C_j)$.
It must execute before or after each stage reservation $[s_k^V,b_k)$:
\begin{equation}
    C_j\leq s_k^V\quad\lor\quad b_k\leq u_j,
    \qquad \forall j\in\mathcal J_{\mathrm{MPPI}},\;\forall k.
\end{equation}
Reductions also execute in disjoint intervals:
\begin{equation}
    C_j\leq u_l\quad\lor\quad C_l\leq u_j,
    \qquad \forall j,l\in\mathcal J_{\mathrm{MPPI}},\;j<l.
\end{equation}
Finally, every MPC invocation must satisfy its control deadline:
\begin{equation}
    C_j\leq d_j,
    \qquad \forall j.
\label{eq:mpc_deadline}
\end{equation}
This constraint enforces the control deadlines under the modeled execution times.
If the admitted jobs cannot be placed within their deadlines using the available stages, the formulation reports the window as infeasible.
A runtime deadline guarantee requires conservative latency bounds covering execution variability.

\section{Experimental Setup}
\label{sec:exp}

\begin{table} 
\caption{Evaluation benchmarks.}
\resizebox{\columnwidth}{!}{
\renewcommand*{\arraystretch}{1}
\renewcommand*{\tabcolsep}{5pt}
\begin{tabular}{ cccc } 
\toprule[0.15em]
\textbf{\specialcell{Algorithm}} & \textbf{Dataset} & \textbf{Inference Paradigm} & \textbf{Venue} \\ 
\midrule[0.05em]
OpenVLA~\cite{kim24openvla} & LIBERO~\cite{liu2023libero} & Autoregressive & CoRL 2024 \\
\midrule[0.05em]
OpenVLA-OFT~\cite{kim2025fine} & LIBERO~\cite{liu2023libero} & Prefill-Only & RSS 2026 \\

\midrule[0.05em]
UVA~\cite{li2025unified} & PushT~\cite{gym-pusht} & Diffusion & RSS 2025 \\
\bottomrule[0.15em]
\end{tabular}
}
\label{tab:sw}
\end{table}

\para{Evaluation Platforms.}
We evaluate \proj on two Nvidia mobile GPU platforms: Jetson AGX Orin~\cite{orinsoc} and Jetson AGX Thor~\cite{thorsoc}.
Jetson AGX Orin has an Ampere mobile GPU with 275~TOPS of peak performance.
Jetson AGX Thor integrates a Blackwell mobile GPU and delivers 2070~TOPS peak performance.
Both platforms integrate Tensor cores and CUDA cores, while Thor provides additional hardware features for warp specialization.
During the measurement, Nvidia Orin and Thor SoCs run at 1.30 and 1.53~GHz, respectively. 
Their power modes are 60~W and 120~W, respectively.
VLA models run at BF16 precision, while MPC algorithms run at FP32 precision.

\para{Algorithms.}
We evaluate \proj with three representative VLA models covering three inference paradigms described in \Sect{sec:bg:vla}: OpenVLA~\cite{kim24openvla}, OpenVLA-OFT~\cite{kim2025fine}, and UVA~\cite{li2025unified}.
OpenVLA uses an autoregressive inference paradigm, OpenVLA-OFT performs PO-based inference, and UVA adopts a diffusion-based inference paradigm.
As shown in \Tbl{tab:sw}, these models represent different VLA execution patterns and provide a comprehensive evaluation.

\begin{figure}[t]
    \centering
    \includegraphics[width=1\columnwidth]{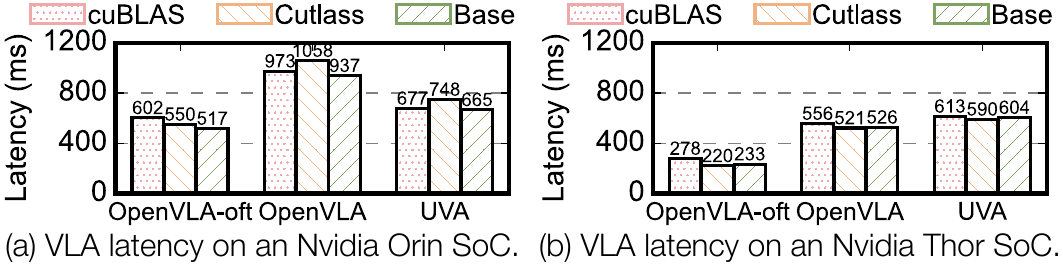}
    \caption{The absolute latency of different VLA models in cuBLAS, CUTLASS, and our implementation.}
    \label{fig:vla_abs_latency}
\end{figure}

\para{Algorithm Implementation.} 
To support the three co-location mechanisms in \Sect{sec:framework:mechanism}, we customized our CUDA kernels.
Although cuBLAS and CUTLASS provide off-the-shelf GEMM implementations, they cannot directly support all the co-location mechanisms required by \proj.
\Fig{fig:vla_abs_latency} compares the inference latency of our implementation against cuBLAS and CUTLASS on Orin and Thor SoCs.
Our implementation achieves comparable performance across VLA models and GPU platforms.
For MPC algorithms, we implement the QP-based and MPPI-based solvers based on the open-source cuOSQP~\cite{schubiger2020gpu} and MPPI-Generic~\cite{vlahov2026mppi} implementations, respectively.

\para{Optimizer Implementation.}
We implement our co-location optimization from \Sect{sec:solver} in Python with about 1k lines of code.
Then, we leverage the Google optimization library \textit{OR-Tools}~\cite{ortool} to solve the constrained optimization problem.

\para{Software Setup.}
We evaluate the VLA models on two robotic datasets: LIBERO~\cite{liu2023libero} and PushT~\cite{gym-pusht}.
For closed-loop execution, we integrate MPC with the MuJoCo simulator~\cite{6386109}, where the optimized control commands are applied to the simulated robot and the updated states are fed back to MPC.
VLA and MPC execute asynchronously, with MPC using the latest available VLA-generated reference.
We report three metrics: (1) end-to-end latency of each VLA-MPC control cycle, (2) energy consumption during execution, and (3) MPC QoS hit rate, defined as the fraction of MPC jobs whose control results are available by their deadlines. 
Unless otherwise specified, MPC runs at a target frequency of 500~Hz~\cite{neuman2021robomorphic}, and each MPC job is due 2~ms after its periodic release.
Here, we report speedup relative to the sequential execution baseline.
The co-location latency includes waiting for the completion of the outstanding MPC jobs and the final reduction time in MPPI.
We also compute GPU energy by integrating readings from the built-in GPU power sensor over the test interval.

\para{Baselines.}
Since there is no prior work targeting VLA-MPC co-location directly, we implement several baselines based on prior work.
We compare \proj with two sequential baselines and four single-mechanism co-location baselines.
For a fair comparison, all four co-location baselines use the optimizer in \Sect{sec:solver} to determine their execution configurations and MPC-to-VLA mappings, but are restricted to a single co-location mechanism.
In contrast, Dubot considers all available mechanisms and selects the most suitable configuration for each VLA stage.
\begin{itemize}
    \item \mode{CPU}: This baseline executes VLA inference on the GPU and MPC computation on the CPU of the same SoC. Here, VLA and MPC are executed sequentially.
    \item \mode{Base}: This baseline executes VLA inference and MPC computation sequentially on the GPU with no co-location.
    \item \mode{Spatial}: This baseline implements a spatial partitioning approach similar to prior work~\cite{wu2015enabling} by partitioning the GPU at the SM level and assigning different SM groups to VLA inference and MPC computation.
    \item \mode{Stream}: This baseline adopts block-level co-location by executing VLA and MPC as separate kernels in different CUDA streams. Their thread blocks can share an SM when sufficient resources are available.
    \item \mode{WASP-Limited}: This baseline performs intra-block fusion by partitioning threads within a block into different groups, similar to prior work, Tacker~\cite{zhao2022tacker}. We only report its performance on Orin in our evaluation.
    \item \mode{WASP}: 
    This approach uses warp specialization~\cite{crago2024wasp} to assign different warps within a block to dedicated tasks. It also supports register reallocation (\texttt{setmaxnreg})~\cite{pts_doc} and asynchronous Tensor core execution~\cite{blackwell_doc}. Note that these features are only supported by the Nvidia Thor SoC; we only report its performance on Thor in our evaluation.
    \item \mode{\proj}: Our proposed framework considers all three co-location mechanisms. It automatically selects the most suitable mechanism for each VLA stage using our optimizer.
\end{itemize}
\section{Evaluation}
\label{sec:eval}

\subsection{Performance}
\label{sec:eval:perf}

\begin{figure}[t]
    \centering
    \includegraphics[width=1\columnwidth]{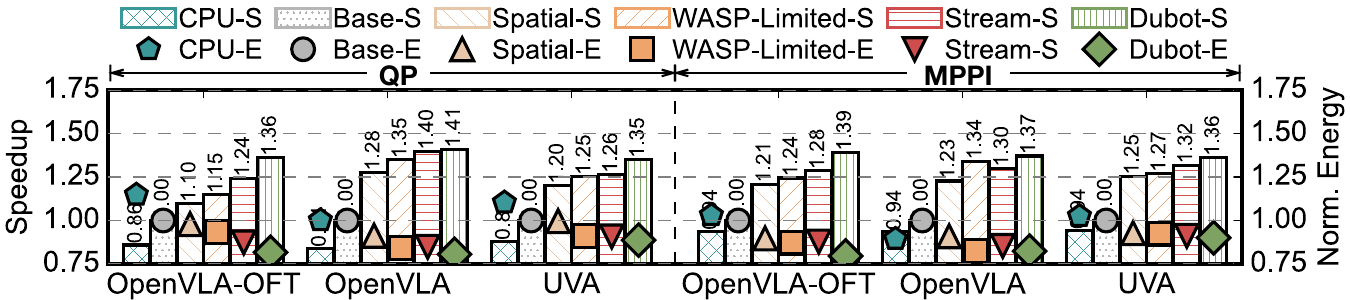}
    \caption{Comparison of performance and normalized energy among different baselines on an Orin SoC. Numbers are normalized to \mode{Base}. `S': speedup; `E': normalized energy.}
    \label{fig:orin_speedup_energy}
\end{figure}

\para{Performance on Orin.}
\Fig{fig:orin_speedup_energy} compares the performance and energy savings of \mode{\proj} against different baselines on an Nvidia Orin SoC.
All results are normalized to \mode{Base}, i.e., the GPU sequential execution baseline.
Compared with \mode{Base}, \mode{Spatial} provides limited speedup (1.21$\times$) because it maps VLA and MPC computations onto different SM groups and cannot simultaneously utilize Tensor cores and CUDA cores within the same SM.

Intra-block fusion approaches, e.g., \mode{WASP-Limited}, can better overlap the computation between the Tensor cores and CUDA cores.
On average, \mode{WASP-Limited} achieves 1.27$\times$ speedup.
However, \mode{WASP-Limited} on Orin still suffers from the extra overhead due to limited hardware support for warp specialization.

\mode{\proj} achieves the highest speedup across different EAI pipelines, with up to 1.41$\times$ speedup.
This is because \proj can choose different co-location mechanisms for different operators.
Moreover, \mode{\proj} achieves the highest energy efficiency among the others.
\mode{\proj} reduces energy consumption by 16.1\% compared with \mode{Base}.

\begin{figure}[t]
    \centering
    \includegraphics[width=1\columnwidth]{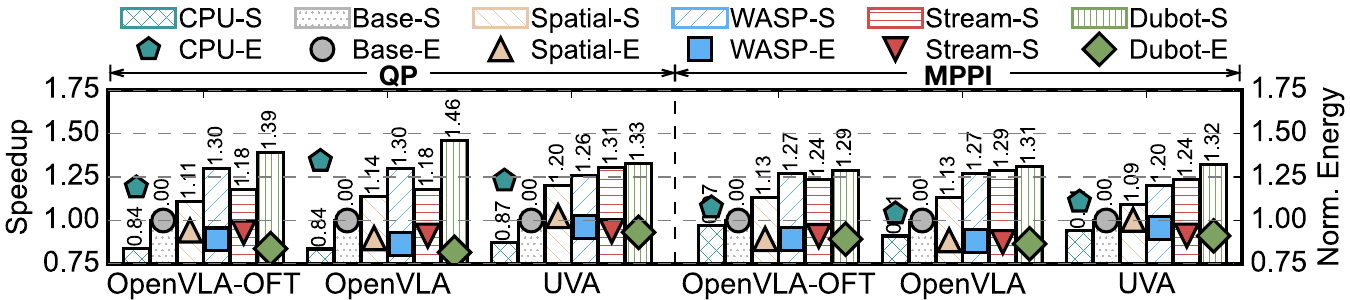}
    \caption{Comparison of performance and normalized energy among different baselines on a Thor SoC. Numbers are normalized to \mode{Base}. `S': speedup; `E': normalized energy.}
    \label{fig:thor_speedup_energy}
\end{figure}

\para{Performance on Thor.}
\Fig{fig:thor_speedup_energy} compares the performance and energy efficiency of \mode{\proj} with various baselines on the Thor SoC.
All performance results are normalized to \mode{Base}.
Here, we include an additional baseline, \mode{WASP}, which exploits newer architectural support for warp-specialized execution.
Overall, \mode{WASP} achieves the second-highest performance, 1.27$\times$ speedup over \mode{Base}.
Nevertheless, \mode{\proj} still outperforms other baselines with 1.35$\times$ speedup.
For energy savings, \mode{\proj} reduces GPU energy by 12.4\% on Thor.

\begin{figure}[t]
    \centering
    \includegraphics[width=1\columnwidth]{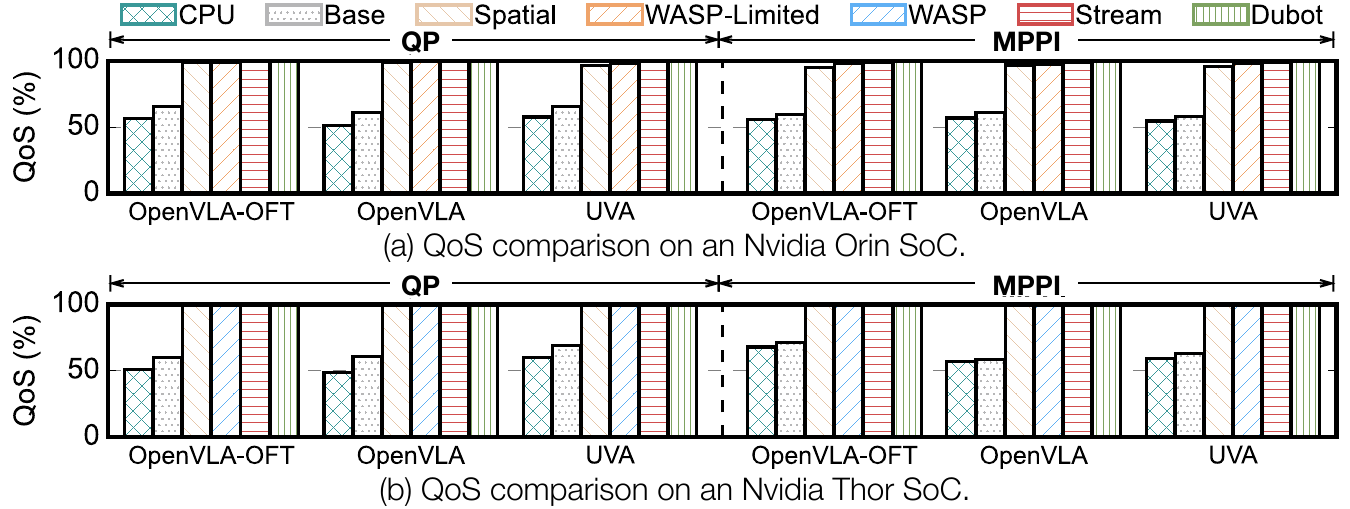}
    \caption{The QoS Comparison among different baselines on two mobile SoCs. The default MPC frequency is 500~Hz.}
    \label{fig:qos_comparison}
\end{figure}

\begin{figure}[t]
    \centering
    \includegraphics[width=1\columnwidth]{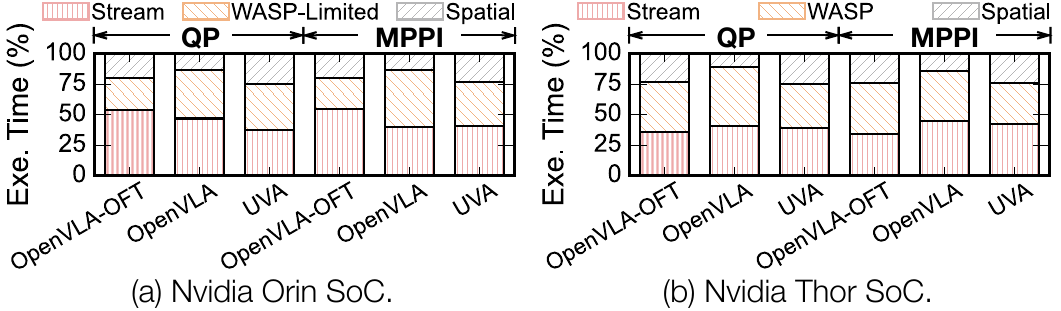}
    \caption{The breakdown of co-location mechanisms selected by \proj on two mobile SoCs.}
    \label{fig:mechanism_breakdown}
\end{figure}

\para{QoS.}
\Fig{fig:qos_comparison} compares the MPC QoS achieved by different baselines on two Nvidia SoCs, Orin and Thor, at a target control frequency of 500~Hz.
Because \mode{Base} sequentially executes the VLA inference and MPC computations, \mode{Base} frequently delays MPC execution behind VLA inference, achieving only 61.7\% of QoS on Orin and 63.7\% on Thor, on average.
In contrast, all co-location baselines largely improve QoS by overlapping the two workloads.
On both Orin and Thor SoCs, \mode{\proj} achieves the best QoS, with over 99\% of QoS across all VLA models and MPC algorithms.

\para{Mechanism Breakdown.}
\Fig{fig:mechanism_breakdown} shows the breakdown of co-location mechanisms selected by \proj across different VLAs and MPCs.
On Orin, \textit{Stream} and \textit{WASP-Limited} account for 45.8\% and 35.3\% of the selections on average, respectively. 
On Thor, both \textit{Stream} and \textit{WASP} are frequently selected, with WASP benefiting from additional hardware support for warp specialization.
Thus, different VLA operators favor different co-location mechanisms.

\begin{figure}[t]
    \centering
    \includegraphics[width=1\columnwidth]{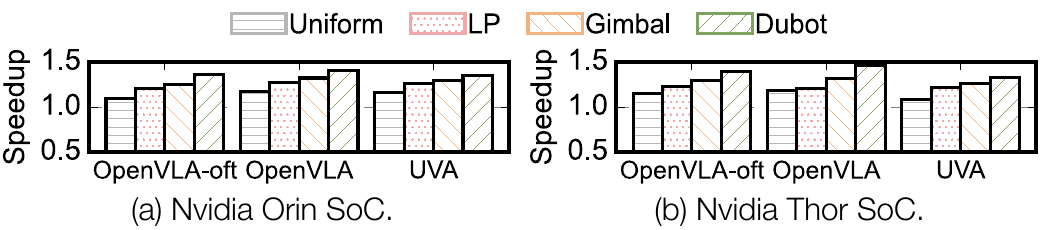}
    \caption{Ablation study on our co-location optimizer.}
    \label{fig:optimizer_ablation}
\end{figure}

\subsection{Ablation Study}
\label{sec:eval:abl}

In this subsection, we evaluate the contribution of our optimizer by replacing it with three heuristics.
\mode{Uniform} distributes QP iterations as evenly as possible across VLA operators.
\textit{Latency Proportional} (\mode{LP}) assigns iterations in proportion to each operator's standalone latency.
\mode{Gimbal} is a greedy co-location algorithm~\cite{yang2025taming}.
Each baseline can select the best-performing co-location mechanism for its fixed QP assignment.
Here, we only evaluate the QP-based MPC algorithm over three VLA models.

In \Fig{fig:optimizer_ablation}, \mode{LP} consistently outperforms \mode{Uniform}.
This shows that standalone VLA latency is a useful signal for workload placement.
Nevertheless, \proj achieves the highest speedup for all three VLA models on both Jetson Orin and Thor, even compared with \mode{Gimbal}.
\mode{\proj} speedup ranges from $1.33\times$ to $1.46\times$.
Overall, \proj outperforms all heuristic methods across configurations.

\subsection{Sensitivity Study}
\label{sec:eval:sens}

\begin{figure}[t]
    \centering
    \includegraphics[width=1\columnwidth]{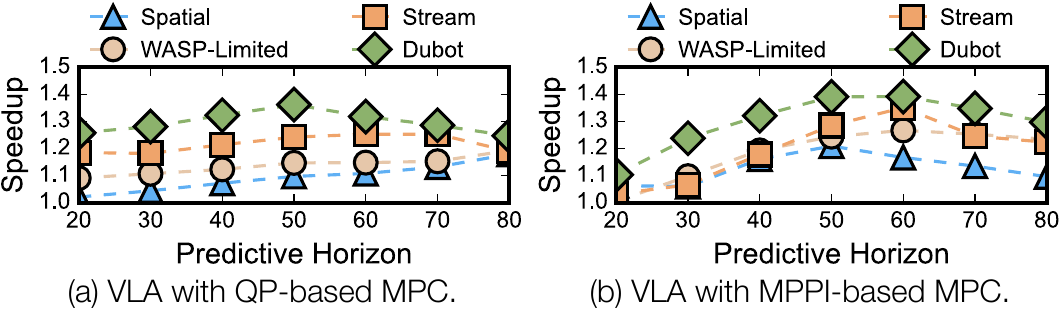}
    \caption{The sensitivity of performance to predictive horizon with two MPC solvers, QP and MPPI, on an Orin SoC. The default MPC frequency is 500~Hz.}
    \label{fig:predictive_horizon_sens}
\end{figure}

\para{Predictive Horizon.}
\Fig{fig:predictive_horizon_sens} shows the sensitivity of speedup to the MPC predictive horizon on the Orin SoC with a target frequency of 500~Hz.
The length of the MPC predictive horizon is proportional to the MPC workload.
Here, we choose OpenVLA-OFT as a case study.
For both QP-based and MPPI-based MPC, as the predictive horizon increases, the performance of all co-location strategies initially improves.
This is because a larger MPC workload exposes more opportunities to overlap with VLA inference.
However, the speedup eventually saturates and slightly decreases at larger horizon windows as the MPC workload begins to exceed the overlap window and dominates the execution latency.
Both \mode{Spatial} and \mode{WASP-Limited} show moderate speedup.
Overall, \mode{\proj} achieves 1.3$\times$ average speedup when the horizon is between 30 and 70.

\begin{figure}[t]
    \centering
    \includegraphics[width=1\columnwidth]{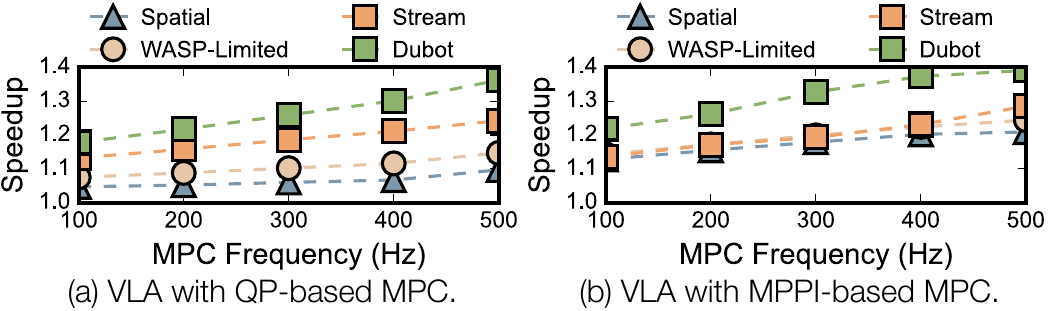}
    \caption{The sensitivity of performance to MPC frequency with two types of MPC solvers, QP-based and MPPI-based, on Orin SoC. The default predictive horizon length is 50. The speedup is normalized to the corresponding \mode{GPU} baseline.}
    \label{fig:mpc_fre_sens}
\end{figure}

\para{MPC Target Frequency.}
\Fig{fig:mpc_fre_sens}a and \Fig{fig:mpc_fre_sens}b show the sensitivity of speedup to the MPC target frequency on an Orin SoC, using two MPC solvers: QP-based and MPPI-based.
The default predictive horizon length is 50.
All numbers are normalized to the \mode{GPU} baseline.
Here, we also use OpenVLA-OFT.
For the QP-based solver (\Fig{fig:mpc_fre_sens}a), all co-location approaches maintain a stable increase in speedup in the frequency range from 100 to 500~Hz.
Overall, \mode{\proj} achieves roughly 1.3$\times$ speedup on average.
For MPPI, as the frequency increases, the trend is similar.

\subsection{Case Study}
\label{sec:eval:case}


\begin{figure}[t]
    \centering
    \includegraphics[width=\columnwidth]{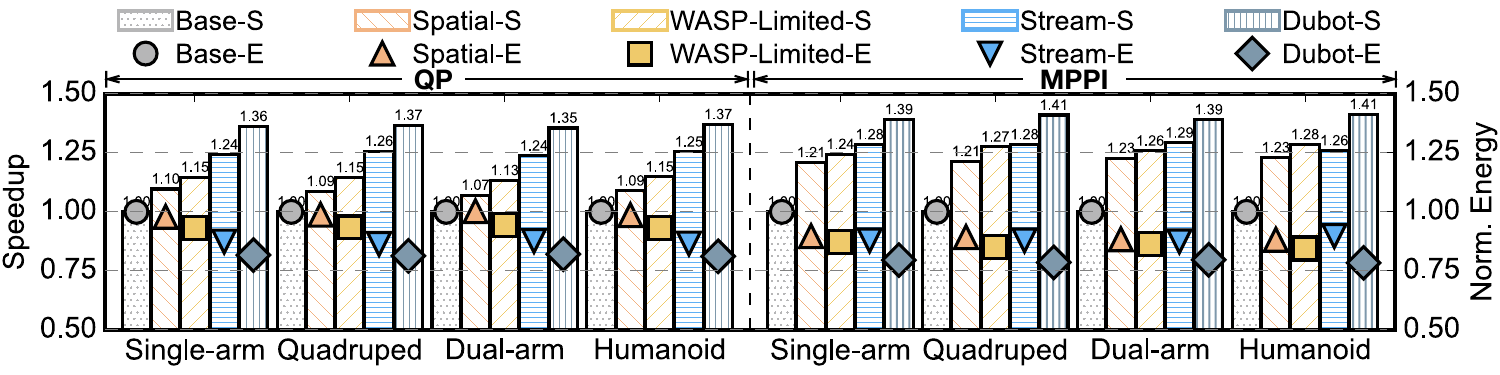}
    \caption{Scalability study of \proj with two different MPC solvers, QP-based and MPPI-based, on an Orin SoC. Here, we show four different robot types: a single-arm manipulator, a quadruped robot, a dual-arm manipulator, and a humanoid.
    Speedup is normalized to the corresponding \mode{GPU} baseline.}
    \label{fig:scalability_study}
\end{figure}

To evaluate the generality of \proj, we measure its performance across four representative robot platforms: a single-arm manipulator, a quadruped robot, a dual-arm robot, and a humanoid.
To mimic the workloads of different robot platforms, we fix the workload of the VLA model and use OpenVLA-OFT as a case study.
We then emulate MPC workloads derived from four robot configurations.
These robot configurations are derived from the reported specifications and degrees of freedom of the platforms~\cite{khazatsky2024droid, he2024agile, huang2025prrtc, xialin2025learning}.

\Fig{fig:scalability_study}a and \Fig{fig:scalability_study}b show the speedup and normalized energy of different co-location variants on the Orin SoC with two MPC solvers, QP-based and MPPI-based.
The performance numbers are normalized to the corresponding \mode{GPU} baselines.
Across all four robot types, \mode{\proj} consistently achieves the highest speedup, roughly 1.4$\times$.
The speedup of \mode{\proj} remains stable because \mode{\proj}'s co-location optimizer automatically adapts the mapping of the available GPU resources between VLA kernels and MPC computations.

For energy savings, \mode{\proj} also achieves the best energy efficiency across all robot platforms.
It reduces energy consumption by roughly 20\% compared to \mode{Base}.
On the other hand, \mode{Spatial} provides only modest energy improvement, as statically splitting SMs between the two workloads can lead to idle compute units on both partitions.
Overall, the results show that \mode{\proj} generalizes effectively across diverse robot types without manual tuning.

\section{Related Work}
\label{sec:related}

\para{MPC Acceleration.}
MPC is a key component in robotic systems and has received extensive attention from the architecture community~\cite{sacks2018robox, wang2023rsqp, hao2024orianna, ting2025hiper, neuman2021robomorphic, neuman2023roboshape, bakhshalipour2024tartan, yang2023dadu}.
For instance, RoboX~\cite{sacks2018robox} targets the conventional robotic MPC pipeline, while RSQP~\cite{wang2023rsqp} accelerates its underlying quadratic programs.
Others~\cite{hao2024orianna, ting2025hiper} target optimization-based robotic workloads. 
Robomorphic~\cite{neuman2021robomorphic}, RoboShape~\cite{neuman2023roboshape}, and Dadu-RBD~\cite{yang2023dadu} accelerate rigid-body dynamics, whereas Tartan~\cite{bakhshalipour2024tartan} designs a programmable processor for robotic control.

However, existing techniques primarily treat MPC as an isolated workload and optimize its efficiency.
\proj explores the computational characteristics between MPC and VLA, and designs a co-location framework on off-the-shelf GPUs.

\para{VLA Acceleration.}
Recent VLA acceleration methods mainly focus on algorithmic optimization.
Some~\cite{wen2024tinyvla, liurobomamba, zhang2026quantvla, zheng2026dyqvla} reduce inference cost through compression and quantization techniques. 
Others~\cite{zhang2025mole, yang2026dysl} dynamically skip unnecessary layers or decoding steps for similar inputs. 
Token-pruning approaches~\cite{xu2025vla, tan2025thinktwiceactonce, wang2026specprune} remove redundant visual tokens while preserving task-level precision.
Recently, there have been some algorithm-hardware co-designs for VLA~\cite{liu2026deltoris, qi2026specvla, song2026vqvla}.
Different from these approaches that reduce the cost of VLA inference alone, \proj focuses on system-level optimization to improve end-to-end EAI performance on mobile GPUs.

\para{Kernel Fusion and Co-Location.}
Recent kernel fusion techniques reduce execution overhead by combining operators, minimizing intermediate memory traffic and kernel launches~\cite{zhao2023ispa, yang2025taming, zhao2022tacker, zhao2025adaptive}. 
Some~\cite{chen2018tvm, zheng2020ansor, zhu2022roller} enable compiler-driven fusion via tensor scheduling. 
Others propose various techniques to manage interference between GPU workloads~\cite{han2022reef, strati2024orion}.
Further, studies~\cite{kao2023flat, hu2024optimal, qin2025chimera, park2026trinity} explore domain-specific optimization and communication. 
In contrast, \proj performs robotics-specific optimization via co-location.

\para{Autonomous System Acceleration.}
Recent architectural research has explored a broad range of techniques to accelerate robotic systems~\cite{liu2021micro, krishnan2023autopilot, shah2023energy, shah2024collision, choe2025ana, huang2025corki, chen2025octocache, wan2025reca, huang2026splatonic, li2025rtgs}.
Some mainly focus on optimizing individual components in robotics, such as localization~\cite{liu2021micro, huang2026splatonic, li2025rtgs} and collision detection~\cite{shah2023energy, shah2024collision}.
Recent architectural efforts have begun targeting embodied AI and autonomous driving~\cite{huang2025corki, wan2025reca, chen2025octocache, choe2025ana}.
\proj also speeds up the VLA-MPC loops via co-optimizing its algorithms holistically.


\section{Conclusion}

Future EAI will not be defined solely by larger foundation models, but by how effectively the high-level planning loop and the low-level control loop are integrated and executed together.
This integration also calls for a new generation of EAI architectures that jointly optimize model design, runtime system, and architecture.


\bibliographystyle{ACM-Reference-Format}
\bibliography{refs}

\end{document}